\documentclass[a4paper,11pt]{article}
\pdfoutput=1 

\usepackage{jheppub} 
\usepackage[bottom]{footmisc}
\usepackage{amssymb}
\usepackage{amsmath}
\usepackage{amsthm}
\usepackage[usenames,dvipsnames]{xcolor}
\usepackage{epsfig}
\usepackage{dcolumn}
\usepackage{tikz}
\usetikzlibrary{shapes.geometric, arrows}
\usepackage{upgreek}
\usepackage{setspace}
\usepackage{enumitem}
\usepackage{array,multirow,bigdelim,arydshln}
\usepackage{appendix}
\usepackage{xparse}
\usepackage{stmaryrd}
\usepackage[T1]{fontenc} 
\usepackage{mathtools}
\usepackage{physics} 
\usepackage{adjustbox}
\usepackage{multirow}
\usepackage{graphicx} 
\usepackage{float} 
\graphicspath{{./images/}}
\usepackage[nottoc]{tocbibind}
\usepackage{hyperref}
\usepackage[utf8]{inputenc}
\hypersetup{
	colorlinks,
	urlcolor=Maroon,
	linkcolor=Maroon,
	citecolor=Maroon
	}

\NewDocumentCommand{\binomial}{omm}
 {%
  \genfrac(){0pt}{}{#2}{#3}%
  \IfValueT{#1}{_{\!#1}}%
 }
\NewDocumentCommand{\eulerian}{omm}
 {%
  \genfrac<>{0pt}{}{#2}{#3}%
  \IfValueT{#1}{_{\!#1}}%
 }

\def \s {\sigma}

\usepackage{latexsym}
\usepackage{tikz}

\theoremstyle{plain}

\theoremstyle{definition}
\newtheorem{definition}{Definition}[section]

\newtheorem{conjecture}{Conjecture}[section]

\newcommand{\sfs}{\mathsf{s}}
\newcommand{\sft}{\mathsf{t}}
\newcommand{\sfR}{\mathsf{R}}
\newcommand{\sfW}{\mathsf{W}}
\newcommand{\sfU}{\mathsf{U}}

\title{Generalized Planar Feynman Diagrams: Collections}

\author[a,b]{Francisco Borges}\emailAdd{fborges@pitp.ca}
\author[a]{and Freddy Cachazo}\emailAdd{fcachazo@pitp.ca}

\affiliation[a]{Perimeter Institute for Theoretical Physics, Waterloo, ON N2L 2Y5, Canada}
\affiliation[b]{Department of Physics $\&$ Astronomy, University of Waterloo, Waterloo, ON N2L 3G1, Canada}

\abstract{Tree-level Feynman diagrams in a cubic scalar theory can be given a metric such that each edge has a length. The space of metric trees is made out of orthants joined where a tree degenerates. Here we restrict to planar trees since each degeneration of a tree leads to a single planar neighbor. Amplitudes are computed as an integral over the space of metrics where edge lengths are Schwinger parameters. In this work we propose that a natural generalization of Feynman diagrams is provided by what are known as metric tree arrangements. These are collections of metric trees subject to a compatibility condition on the metrics. We introduce the notion of \emph{planar collections of Feynman diagrams} and argue that using planarity one can generate all planar collections starting from any one. Moreover, we identify a canonical initial collection for all $n$. Generalized $k=3$ biadjoint amplitudes, introduced by Early, Guevara, Mizera, and one of the authors, are easily computed as an integral over the space of metrics of planar collections of Feynman diagrams.}

\begin{document}
\maketitle
\addtocontents{toc}{\protect\setcounter{tocdepth}{1}}
\def \tr {\nonumber\\}
\def \la  {\langle}
\def \ra {\rangle}
\def\hset{\texttt{h}}
\def\gset{\texttt{g}}
\def\sset{\texttt{s}}
\def \be {\begin{equation}}
\def \ee {\end{equation}}
\def \ba {\begin{eqnarray}}
\def \ea {\end{eqnarray}}
\def \k {\kappa}
\def \h {\hbar}
\def \r {\rho}
\def \l {\lambda}
\def \be {\begin{equation}}
\def \en {\end{equation}}
\def \bes {\begin{eqnarray}}
\def \ens {\end{eqnarray}}
\def \red {\color{Maroon}}
\def \pt {{\rm PT}}
\def \s {\sigma} 
\def \ls {{\rm LS}}
\def \ma {\Upsilon}
\def \s {\textsf{s}}
\def \t {\textsf{t}}
\def \R {\textsf{R}}
\def \W {\textsf{W}}
\def \U {\textsf{U}}
\def \e {\textsf{e}}

\numberwithin{equation}{section}

\section{Feynman Diagrams as Metric Trees}

Tree-level scattering amplitudes in a cubic scalar field theory have a Feynman diagram expansion in terms of trees with $n$ leaves and $n-2$ trivalent vertices. Adding a $U(N)\times U(\tilde N)$ flavor structure in the biadjoint representation allows the definition of double partial ordered amplitudes traditionally denoted as $m_n(\alpha,\beta)$ (see e.g. \cite{Cachazo:2013iea}). Here $\alpha$ and $\beta$ are two planar orderings and the amplitude is given by a sum over Feynman diagrams that are planar under both orderings. In this work we mainly consider $m_n(\mathbb{I},\mathbb{I})$, where $\mathbb{I}$ is the canonical order $\qty{1,2,\ldots,n-1,n}$, and therefore we will simply refer to the relevant diagrams as planar.

Feynman diagrams can be given a metric. A metric is a symmetric $n\times n$ matrix with non-negative entries, $d_{ab}$, defining the minimum distance between leaves $a$ and $b$, such that it can be obtained from some assignment of non-negative edge lengths, including edges with a leaf. In physics, lengths for internal edges are nothing but Schwinger parameters. However, their main application has traditionally been at loop-level (see e.g \cite{Strassler:1992zr}). In other areas the space of metric trees has been important and carefully studied, for example when metric trees are identified with phylogenetic trees \cite{BilleraL}.  

In physical applications only the contribution of a tree $T$ to an amplitude is of importance. This is easily obtained by first constructing the function
\be\label{metricT}
F(T) := \sum_{1\leq a,b\leq n} d_{ab}\, s_{ab},
\ee
where $s_{ab}$ are the standard $n$-particle Mandelstam kinematic invariants satisfying
\be\label{condT}
s_{ab}=s_{ba}, \quad s_{aa}=0, \quad  \sum_{b=1}^n s_{ab} =0 \quad  \forall a.
\ee
Let us denote the length of the edges containing a leaf as $e_a$ and therefore $d_{ab} = e_a +e_b +d^{\rm int.}_{ab}$, where $d^{\rm int.}_{ab}$ is the contribution from internal edges. It is easy to see that due to momentum conservation the external edges $e_a$ drop out of $F(T)$. Denoting by $f_I$ the length of internal edges one finds from \eqref{metricT}
\be
F(T)= \sum_{1\leq a,b\leq n} d^{\rm int.}_{ab}\, s_{ab} = -\sum_{I=1}^{n-3}f_I\, t_I,
\ee
where $t_I$ is a linear combination of $s_{ab}$ obtained as follows. Deleting an internal edge divides ${\cal T}$ into two trees and hence the set of leaves into two. Denoting the sets as $L_I$ and $R_I$, the coefficient of $f_I$ is clearly
\be
t_I =\, -\!\!\!\!\!\!\sum_{a\in L_I ,b\in R_I}\!\!\!s_{ab}
\ee
since $f_I$ contributes to all $d_{ab}$ for which $a\in L_I$ and $b\in R_I$. Moreover, this is nothing but the inverse of the standard Feynman propagator associated to the internal edge, after using momentum conservation. The contribution to an amplitude is then obtained by multiplying all $n-3$ propagators\footnote{Up to trivial factors of $2$ which can be absorbed in the definition of the coupling constant.}. Using the Schwinger parametrization of the propagators, identifying Schwinger parameters to the lengths, one can write the contribution to an amplitude in term of $F(T)$ as follows
\be\label{intFT}
\frac{1}{\prod_{I=1}^{n-3}t_I} = \prod_{I=1}^{n-3}\int_0^\infty \dd{f_I} \, {\rm exp} \left(-t_I f_I\right) = \int_{(\mathbb{R}^+)^{n-3}}d^{n-3}f_I \, {\rm exp}\, F(T).
\ee

In this work we propose that a natural generalization of this construction is given by what are known as metric tree arrangements as defined by Herrmann, Jensen, Joswig, and Sturmfels \cite{herrmann2009draw}. Moreover, they lead to the $k=3$ generalized biadjoint amplitudes recently introduced by Early, Guevara, Mizera, and one of the authors \cite{Cachazo:2019ngv} and further studied in \cite{Cachazo:2019apa,Drummond:2019qjk,Sepulveda:2019vrz}. 

Before discussing metric tree arrangements, let us discuss the important restriction of planarity mentioned above. In order to compute $m_n(\mathbb{I},\mathbb{I})$ one has to determine all planar Feynman diagrams, i.e., all labeled tree diagrams that can be drawn on a disk without crossing and with the leaves on the boundary of the disk with order $\mathbb{I}=\{1,2,\ldots,n\}$.

Let us start with a planar tree $T$. The lengths of the internal edges $f_I$ provide coordinates on the space of trees. The tree $T$ has $n-3$ degenerations corresponding to collapsing any one of its $n-3$ internal edges, i.e., setting one $f_I=0$. Any degeneration can be resolved in exactly two ways which are consistent with the planarity requirement. In physics terminology one can talk about the $s$ and $t$ channels (the $u$ channel is not allowed). One of the two resolutions leads back to $T$ while the other leads to a new tree $T'$. The new tree is also planar and has a metric and a function $F(T')$ from which the corresponding contribution to the amplitude can be obtained. Each tree has $n-3$ neighbors and the space of all planar trees is connected via such transitions. Of course, if we create a graph with $C_{n-2}$ vertices, where $C_m$ is the $m^{\rm th}$ Catalan number\footnote{Recall that $C_m = \tfrac{1}{m+1}{2m \choose m}$.}, and edges connecting two vertices corresponding to trees that are neighbors, then for physical purposes all we need is a Hamiltonian cycle to construct the amplitude.

In this work we show that the condition of planarity is also a very powerful constraint when dealing with metric tree arrangements. In section \ref{sec:planarfeynman} we introduce the notion of planarity for arrangements and call them \emph{planar collections of Feynman diagrams}. We give evidence that all the properties already mentioned for Feynman diagrams as metric trees also hold for planar collections and show how to construct the set of collections which gives rise to $k=3$ generalized biadjoint amplitudes $m_n^{(k=3)}(\mathbb{I},\mathbb{I})$ by starting from any valid collection and traveling across any of its degenerations. As a simple example, we study the $n=6$ case in detail, find all $48$ planar collections with their connections and a Hamiltonian cycle in the graph with $48$ edges. Moreover, we also compute their contributions to $m_6^{(3)}(\mathbb{I},\mathbb{I})$. For $n=7$ we derive and compute in detail the contribution to $m_7^{(3)}(\mathbb{I},\mathbb{I})$ of planar collections that give rise to $7$, $8$, and $9$ propagators. 

Of course, in order to carry out the procedure for any $n$ one needs at least one valid planar collection. Luckily, for any $n$ one can construct exactly $C_{n-2}$ such planar collections as there is a way of producing a collection using any planar Feynman diagram by removing leaves.

We study the planar collections induced by the ``caterpillar'' diagram for any value of $n$ and find all its $2(n-4)$ degenerations explicitly. In section \ref{sec:clusterconnection} we discuss the connection to cluster algebras which has recently been identified as a useful way of computing generalized biadjoint amplitudes by Drummond, Foster, G{\"u}rdogan, and Kalousios \cite{Drummond:2019qjk}. 

We end the paper with a discussion on what kind of objects are needed to go to higher values of $k$. When $k=4$ these turn out to be collections of collections of Feynman diagrams.

\section{Planar Collections of Feynman Diagrams}
\label{sec:planarfeynman}

The space of metric trees with $n$ leaves is closely related to the space of tropical two-planes, i.e., ${\rm Trop}\, G(2,n)$ \cite{speyer2004tropical,speyer2004tropicalM}. The metric $d_{ab}$ defined in terms of edge lengths automatically satisfies the tropical Pl\"{u}cker relations required for $\pi_{ab}:=d_{ab}$ to be in ${\rm Trop}\, G(2,n)$.

A natural question is what kind of objects replaces metric trees for ${\rm Trop}\, G(3,n)$. The answer is a metric tree arrangement \cite{herrmann2009draw}. In order to motivate this, one can start with ${\rm Trop}\, G(3,5)$ which is isomorphic to ${\rm Trop}\, G(2,5)$. This means that whatever generates the tropical Pl\"{u}cker vectors $\pi_{abc}$ for ${\rm Trop}\, G(3,5)$ must somehow be related to the metric trees governing ${\rm Trop}\, G(2,5)$.

\begin{figure}[t]
    \centering
    \includegraphics[width=0.35\linewidth]{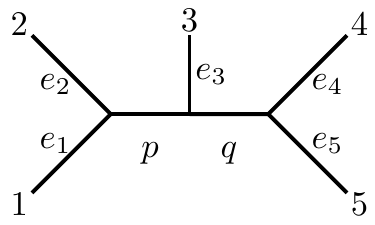}
    \caption{Caterpillar tree with 5 leaves and edge lengths labeled by $e_i$, $p$ and $q$.}
    \label{fig:caterpillarexample}
\end{figure}

Consider the ``caterpillar'' tree with $n=5$ and leaves labeled as in figure \ref{fig:caterpillarexample}. The metric is easily read from the diagram. For example $d_{12}=e_1+e_2$ and $d_{15}=e_1+e_5+p+q$. Note that the internal lengths $p$ and $q$ can be expressed in terms of the metric as follows
\be\label{five}
\begin{aligned}
p &= \frac{1}{4}\left(d_{23}+d_{13}+d_{14}+d_{24}-2d_{12}-2d_{34} \right),\\ \quad q &= \frac{1}{4}\left(d_{24}+d_{25}+d_{34}+d_{35}-2d_{23}-2d_{45} \right).
\end{aligned}
\ee
Moreover the function relating kinematic invariants to the metric can be simplified to be
\be
F(T)  = \sum_{1\leq a,b \leq 5} d_{ab}\, s_{ab} = -4( s_{12}\, p + s_{45}\, q). 
\ee

\begin{figure}[b]
    \centering
    \includegraphics[width=\linewidth]{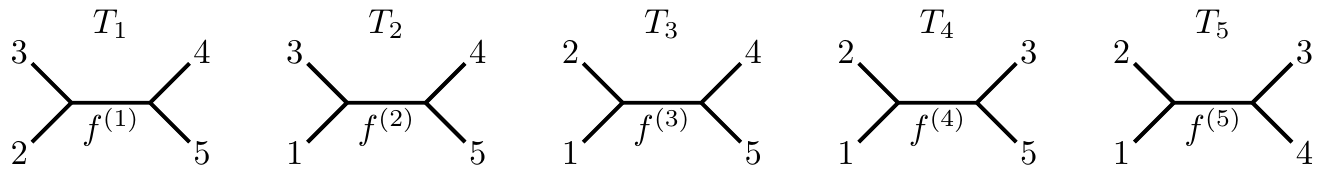}
    \caption{Abstract tree arrangement $\qty{T_1, T_2, T_3, T_4, T_5}$ obtained from the tree in figure \ref{fig:caterpillarexample} by pruning one leaf at the time. It is also a metric tree arrangement.}
    \label{fig:collection5example}
\end{figure}

The object dual to the tree in figure \ref{fig:caterpillarexample} is nothing but a collection of five trees with four leaves obtained from the caterpillar tree by removing one leaf at a time. We show the resulting collection in figure \ref{fig:collection5example}. This is one the simplest examples of what is known as an abstract tree arrangement in the mathematical literature. We review the definition and relation to metric tree arrangements in appendix \ref{ap:treearrs}.

It is natural to give a metric to each tree in the collection of figure \ref{fig:collection5example}. Let the metric of the $a^{\rm th}$ tree be $d^{(a)}_{bc}$. It is instructive to also try and express the internal lengths in terms of the metrics. For example, in the first tree which has leaves $\{2,3,4,5\}$ the internal edge's length is given by
\be
f^{(1)} = \frac{1}{4}\left( d^{(1)}_{35}+d^{(1)}_{34}+d^{(1)}_{25}+d^{(1)}_{24}-2d^{(1)}_{45}-2d^{(1)}_{23}\right).
\ee 
The resemblance with \eqref{five} is an indication of what must be done. Even more so if we compare $f^{(1)}$ with the formula for $q$ in \eqref{five}. Note that if we identify $d^{(1)}_{ij}=d_{kl}$ with $\{k,l\}=\{2,3,4,5\}\setminus \{i,j\}$ then $q=f^{(1)}$. This identification can be made for all five trees in the collection revealing that 
\be\label{comp}
f^{(1)} = q, \; f^{(2)} = q, \; f^{(3)} = p+q,\; f^{(4)} = p, \; f^{(5)} = p. 
\ee  
Moreover, $d^{(a)}_{bc} = d^{(b)}_{ac}=d^{(c)}_{ab}$ and therefore it motivates the introduction of a completely symmetric object $\pi_{abc}:=d^{(a)}_{bc}$. This indeed gives the dual Pl\"{u}cker vector in ${\rm Trop}\, G(3,5)$.

In general one has the following definition.

\begin{definition}\label{metricTree}
\cite{herrmann2009draw} A \emph{metric tree arrangement} on $\{1,2,\ldots ,n\}$ is a collection of $n$ metric trees where the $i^{\rm th}$ tree has $n-1$ leaves given by $\{1,2,\ldots ,n\}\setminus i$ and where the metrics satisfy the compatibility conditions $d^{(a)}_{bc} = d^{(b)}_{ac}=d^{(c)}_{ab}$ for all $a,b,c \in \{1,2,\ldots ,n\}$.
\end{definition}

In appendix \ref{ap:treearrs} we give more details on the definition and relation to ${\rm Trop}\, G(3,n)$.

\subsection{Planarity}

The structure of metric tree arrangements for general $n$ is still an open problem in mathematics \cite{herrmann2009draw}. However, in order to compute $k=3$ generalized amplitudes only a very special class of arrangements are needed. These are the ones that satisfy a planarity condition as explained in the following definition.

\begin{definition}\label{planarC}
A \emph{planar collection of Feynman diagrams} is a metric tree arrangement in which the $i^{\rm th}$ tree is planar with respect to the ordering obtained by deleting the $i^{\rm th}$ label from the canonical order $\mathbb{I} =\{ 1,2,\ldots ,n-1,n\} $.   
\end{definition}

The definition of metric arrangements of planar trees was implicitly given in Remark 4.9 of \cite{herrmann2009draw}. Note that we have chosen to attach the property ``planar'' to the collection and not to individual Feynman diagrams because we think of it is a global property of the collection. This allows us to say that a collection of Feynman diagrams is planar with respect to $\mathbb{I} =\{ 1,2,\ldots ,n-1,n\}$. This is most useful when defining $m_n^{(3)}(\alpha,\beta)$ as in \eqref{qori}.  

One way of constructing all planar collections of Feynman diagrams is the following. Let ${\cal P}_{i,n}$ be the set of planar Feynman diagrams in $n-1$ labels with respect to the ordering obtained by deleting the $i^{\rm th}$ label from the canonical order $\mathbb{I} =\qty{1,2,\ldots ,n-1,n}$. Clearly $|{\cal P}_{i,n}|= C_{n-3}$, with $C_m$ the $m^{\rm th}$ Catalan number. A planar collection of Feynman diagrams is an element of the Cartesian product ${\cal P}_{1,n}\times {\cal P}_{2,n}\times \cdots \times {\cal P}_{n,n}$ which admits a choice of metric for each tree satisfying $d^{(a)}_{bc} = d^{(b)}_{ac}=d^{(c)}_{ab}$. This procedure is clearly not computationally feasible as $n$ becomes large; however, as we explain below this is not necessary.

\subsection{Degenerations}

Consider any planar collection of Feynman diagrams ${\cal T}=\{ T_1,T_2,\ldots ,T_n\}$, i.e. we denote the metric tree in the $i^{\rm th}$ slot as $T_i$. The metric on $T_i$ is $d^{(i)}_{ab}$. In the same way a metric tree is connected to others via degenerations where some internal edge has zero length, collections can also be connected to others. Here is where the power of planarity manifests itself. For each possible degeneration of a planar collection there is a unique neighbor planar collection. 

Let us first discuss degenerations for planar collections before discussing how to use them to construct neighboring planar collections. 

Given a planar collection, ${\cal T}=\{ T_1,T_2,\ldots ,T_n\}$, the naive expectation is that one can degenerate any of the $n-4$ internal edges of any of the $n$ trees in the collection independently thus allowing for $n(n-4)$ degenerations. However, the compatibility conditions on the metrics link some degenerations together and forbid others. This is very explicit in the $n=5$ example introduced above in figure \ref{fig:collection5example}. Let us rewrite the form of the internal edges after the compatibility conditions were imposed \eqref{comp}:
\be\label{comp2}
f^{(1)} = q, \; f^{(2)} = q, \; f^{(3)} = p+q,\; f^{(4)} = p, \; f^{(5)} = p. 
\ee 
Note that sending $f^{(1)}\to 0$ forces $f^{(2)}\to 0$ and thus $f^{(1)}$ and $f^{(2)}$ only give rise to a single degeneration. The same is true for $f^{(4)}$ and $f^{(5)}$. Finally, $f^{(3)}\to 0$ is not valid as a co-dimension one degeneration as this would require either $p$ or $q$ to be negative and therefore it would be out of the space since lengths cannot be negative. Of course, finding that this collection has only two degeneration is welcome as it is supposed to be dual to a single caterpillar metric tree with $n=5$. 

The procedure for finding all possible degenerations of a collection ${\cal T}=\{ T_1,T_2,\ldots ,T_n\}$ is the following. Start by building a $(n-4)\times n$ matrix of the lengths  of all internal edges evaluated on the solution of the compatibility conditions for the metrics and then send each entry, one at a time, to zero to find out the behavior of the whole matrix. Any other entries that become zero define the same degeneration and do not have to be tested. A limit is not valid if it forces us out of the space, i.e., it forces some entries to be negative.

A generic planar collection has $2(n-4)$ degenerations which is surprisingly less than the naive estimate. This shows how strong the compatibility conditions can be. It is important to mention that $2(n-4)$ is the minimum number of degenerations.  

In order to continue with the analogy to metric trees, the next step is generalizing the function $F(T)$ presented in \eqref{metricT} to planar collections of Feynman diagrams. Recall that the compatibility condition turns the collection of metrics into a completely symmetric rank three tensor $\pi_{abc}:=d^{(a)}_{bc}$, therefore it is natural to define 
\be
{\cal F}({\cal T}):= \sum_{1\leq a,b,c\leq n} \pi_{abc}\, \sfs_{abc}.
\ee
Here $\sfs_{abc}$ are the generalized Mandelstam invariants introduced in \cite{Cachazo:2019ngv}. These satisfy conditions analogous to \eqref{condT}
\be\label{condS}
\sfs_{abc}=\sfs_{acb}=\sfs_{cab}, \qquad \sfs_{aab}=0, \qquad  \sum_{b,c=1}^n \sfs_{abc} =0 \quad  \forall a.
\ee

As we show below in an all-multiplicity class of examples and in a variety of $n=6$ and $n=7$ examples, the function ${\cal F}({\cal T})$ only depends on $2(n-4)$ parameters. These can be chosen to be some of the lengths of internal edges of the trees in the collection. 

Since the space of planar collections has exactly $2(n-4)$ parameters, one finds the same behavior as for metric trees, i.e., if one makes some choice of $2(n-4)$ internal edges $f_I$ as coordinates of the space then
\be\label{jumi}
{\cal F}({\cal T}) = -\!\!\!\sum_{I=1}^{2(n-4)}\!\! t_I \, f_I.
\ee 
This time $t_I$ is a linear combination of $\sfs_{abc}$. 

Unlike metric trees, planar collections can have more degenerations than the number of independent parameters and therefore the contribution to a generalized amplitude is not just the product of the $t_I$'s. However, just as for metric trees the contribution is given by an integral over the space of metrics. We postpone the general discussion of amplitudes to section \ref{sec:genamps}. Of course, when the planar collection has exactly $2(n-4)$ degenerations then the contribution to an amplitude is again the inverse of the product of all $t_I$'s in \eqref{jumi}.

\subsection{All-Multiplicity Planar Caterpillar Collection}
\label{sec:allmult}

\begin{figure}[t]
    \centering
    \includegraphics[width=0.5\linewidth]{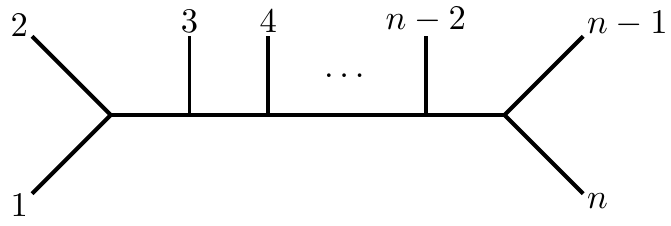}
    \caption{This graph is sometimes called a ``caterpillar'' tree. Leaves $1$ and $n$ are usually called the head and tail but we call the sets $\{1,2\}$ and $\{n-1,n\}$ the cherries of the tree.}
    \label{fig:ncaterpillarexample}
\end{figure}

\begin{figure}[b]
    \centering
    \includegraphics[width=\linewidth]{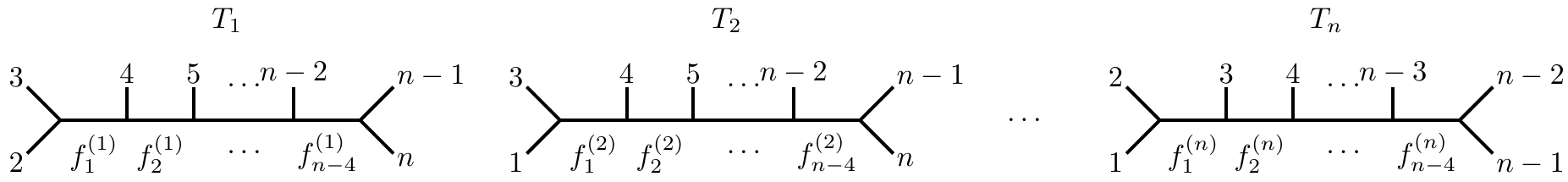}
    \caption{Planar collection obtained from the caterpillar tree with $n$ leaves in figure \ref{fig:ncaterpillarexample}.}
    \label{fig:ncaterpillarcollection}
\end{figure}

Let us illustrate the general procedure with an all $n$ generalization of the $n=5$ caterpillar example discussed above. Here we start with the $n$-point caterpillar tree in figure \ref{fig:ncaterpillarexample} and construct a planar collection of Feynman diagrams by deleting one leaf at a time to get the collection in figure \ref{fig:ncaterpillarcollection}. This construction might seem surprising at first since ${\rm Trop}\, G(2,n)$ is not isomorphic to ${\rm Trop}\, G(3,n)$ for $n\neq 5$. However, as explained in section \ref{sec:higherk}, associating a tree arrangement to a single tree is the first step in the construction of the dual objects associated with ${\rm Trop}\, G(n-2,n)$. 

Going back to the planar collection in figure \ref{fig:ncaterpillarcollection}, note that internal edges have been denoted as $f^{(i)}_\alpha$ with $i\in \{1,2,\ldots ,n\}$ and $\alpha\in \{1,2,\ldots ,n-4\}$. These are the entries of the $(n-4)\times n$ matrix introduced above. 

\begin{figure}[t]
    \centering
    \includegraphics[width=\linewidth]{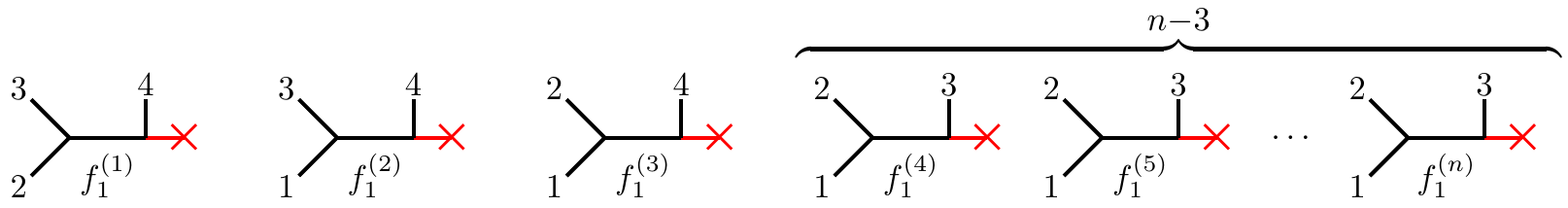}
    \caption{Subtrees defined by the internal edge $f^{(i)}_1$ in the $i^{\rm th}$ tree. The new external edge is represented with a red x.}
    \label{fig:subtrees1}
\end{figure}

At first it seems a daunting exercise to write down all $n$ metrics $d^{(a)}_{bc}$ and impose all compatibility conditions. Luckily, this example has a very simple structure which allows us to reuse the computations done for the $n=5$ case. In order to see this consider the subtrees defined by the internal edge $f^{(i)}_1$ in the $i^{\rm th}$ tree. Separating these as in figure \ref{fig:subtrees1} one notices a replica of the $n=5$ problem with the slight modification that the fourth tree repeats $n-3$ times. This hints that the structure of degenerations in $f^{(i)}_1$ variables is:
\be
f^{(1)}_1=q_1,\; f^{(2)}_1 = q_1, \; f^{(3)}_1 = q_1+ p_1, \; f^{(4)}_1 = p_1,\; f^{(5)}_1 = p_1, \ldots , f^{(n)}_1 = p_1.
\ee  
This shows that out of $n$ internal edges (one per tree in the planar collection), only two degenerations are allowed.

A similar analysis of other edges reveals that for any given $\alpha$, the edge lengths $f^{(i)}_\alpha$ are
\be
f^{(1)}_\alpha=f^{(2)}_\alpha =  \ldots = f^{(\alpha+1)}_\alpha =q_\alpha, \; f^{(\alpha+2)}_\alpha =q_\alpha+ p_\alpha, \; f^{(\alpha+3)}_\alpha = f^{(\alpha+4)}_\alpha = \ldots = f^{(n)}_\alpha = p_\alpha.
\ee  
Since one has exactly $2$ degenerations for every one of the $(n-4)$ rows of the matrix of internal edges, the planar collection associated with the $n$-point caterpillar diagram has $2(n-4)$ degenerations. 

One can complete the analysis by computing the function ${\cal F}({\cal T})$ to find
\be
{\cal F}({\cal T}) = -\!\sum_{i=1}^n\sum_{\alpha=1}^{n-4} t_{i,\alpha}\, f^{(i)}_\alpha.
\ee
A simple computation shows that
\be\label{conte}
{\cal F}({\cal T}) = -\sft_{34\ldots n}\, p_1 - \sum_{i=2}^{n-5}\left(\sft_{i+3\ldots n}\, p_i+\sft_{1\ldots i+1}\, q_i\right) -\sft_{12\ldots n-2}\, q_{n-4}
\ee
where $\sft_{a_1a_2\ldots a_m}$ is a combination of kinematics invariant $\sfs_{abc}$ defined in \cite{Cachazo:2019ngv} and given by
\be
\sft_{a_1a_2\ldots a_m} = \sum_{\{a,b,c\}\subset \{a_1,a_2,\ldots ,a_m\}}\sfs_{abc}.
\ee
As a warm up to the general discussion of amplitudes in section \ref{sec:genamps}, let us compute the contribution of ${\cal T}$ to an amplitudes as this is the simplest case. Note that there are exactly $2(n-4)$ degenerations and therefore the space of metrics is simply $(\mathbb{R}^+)^{2(n-4)}$. The contribution to an amplitude is
\be\label{allN}
{\cal R}({\cal T}) = \int_{({\cal R}^+)^{2(n-4)}}\dd[n-4]{p}\, \dd[n-4]{q}\, {\rm exp}\,{\cal F}({\cal T}) = \frac{1}{\sft_{34\ldots n}\left(\prod_{i=2}^{n-5}\sft_{i+3\ldots n}\sft_{1\ldots i+1}\right)\sft_{12\ldots n-2}}.
\ee

\subsection{Generating All Planar Collections via Planar Moves} 
 
Up to this point the only examples of planar collections we have considered are the ones obtained from the caterpillar tree by removing leaves. It turns out that planar collections can be obtained by the same ``pruning'' operations done on any of the $C_{n-2}$ planar trees in $n$ labels. However, the space of planar collections is much larger. In this section we explain how to generate new planar collections from known ones by a planar move. Although we do not have a proof that every possible planar collection can be reached in this way, we provide many examples in section \ref{sec:examples}. Let us start with formalizing the notion of a planar move.

\begin{definition}\label{planarMove}
Given any planar tree, a \emph{planar move} is a contraction of one of its edges and a re-expansion such that the new tree generated is still planar. Given a planar collection of Feynman diagrams, a planar move is a valid single degeneration of internal edges and a re-expansion of edges that were contracted such that all new trees generated in the new collection are planar. 
\end{definition}

\begin{figure}[t]
    \centering
    \includegraphics[width=0.75\linewidth]{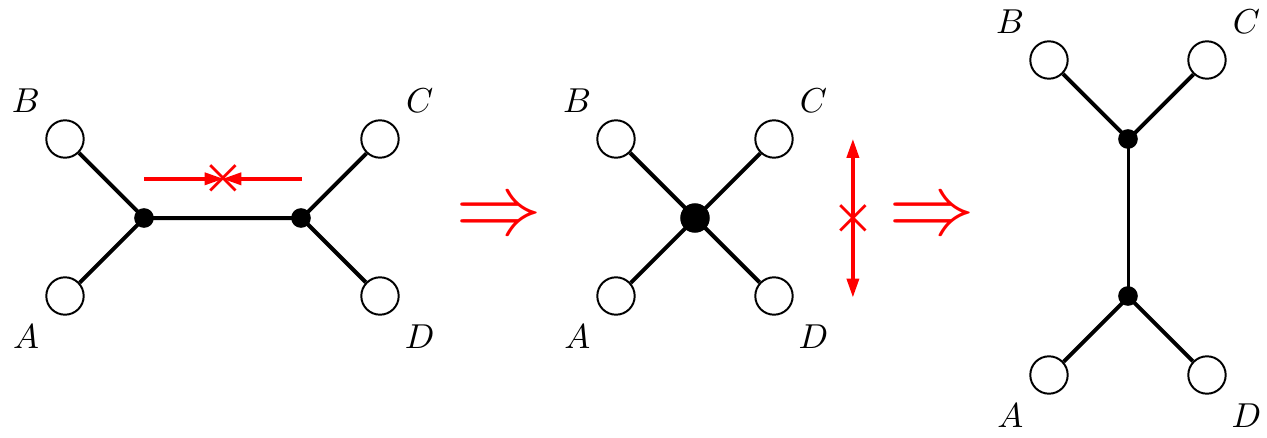}
    \caption{Planar move operation used in Definition \ref{planarMove}.}
    \label{fig:planarmove}
\end{figure}

We illustrate the definition in figure \ref{fig:planarmove}. Note that relaxing the planarity condition allows for another expansion. In physics terms one would say that a planar move connects the $s$-channel to the $t$-channel while the non-planar one connects to the $u$-channel. The extra possibility is not a major source of complication when dealing with the space of metric trees but when metric tree arrangements are considered then the number of possibilities grows very fast. However, the proliferation of possibilities is not main complication, unfortunately most of the new possibilities will not admit a metric and therefore will not be metric tree arrangements. Avoiding this complication is where the main advantage of restricting to planar collections relies as expressed in the following conjecture.

\begin{conjecture}
The set of all $n$-particle planar collections of Feynman diagrams is connected by planar moves. Moreover, given a planar collection of Feynman diagrams, any of its degenerations leads to a neighbor which is also a planar collection of Feynman diagrams.
\end{conjecture}

One of the reasons the conjecture is non trivial is that one has to show that after the unique planar resolution, the new set of trees admits a metric that satisfies the compatibility conditions. Such a proof requires the use of ``non-local'' information, i.e., one has to know the behaviour of the whole collection of Feynman diagrams and not only that of the trees being degenerated.

\subsection{Computing Generalized Amplitudes}
\label{sec:genamps}

Scattering amplitudes in a biadjoint cubic scalar theory can also be computed, in addition to adding Feynman diagrams, by the Cachazo-He-Yuan (CHY) formula \cite{Cachazo:2013hca,Cachazo:2013iea,Cachazo:2014xea}. The CHY formula expresses $m_n(\mathbb{I},\mathbb{I})$ as an integral over the configuration space of $n$ points on $\mathbb{CP}^1$, sometimes denoted as $X(2,n)$. In \cite{Cachazo:2019ngv}, a natural generalization to the configuration space of $n$ points on $\mathbb{CP}^{k-1}$ or $X(k,n)$ with $k>2$ was introduced. Moreover, the corresponding generalized biadjoint amplitude $m_n^{(3)}(\mathbb{I},\mathbb{I})$ was computed for $n=6$ and a proposal for generalized Feynman diagram was given as the facets of ${\rm Trop}\, G(3,6)$. As it tuns out, facets of ${\rm Trop}\, G(3,n)$ are related to metric tree arrangements as reviewed in appendix \ref{ap:treearrs}. 

Here we show that, indeed, planar collections of Feynman diagrams are the natural generalization of Feynman diagrams in that amplitudes in both cases are computed as an integral over the space of metrics.

We have seen already that when a planar collection has exactly $2(n-4)$ degenerations then the space of metrics in it is very simple and given by the $2(n-4)$-dimensional positive orthant of $\mathbb{R}^{2(n-4)}$. We provided an explicit all $n$ example in section 2.3.

The value of the planar collection is obtained by using ${\cal F}(\cal T)$ as written in \eqref{jumi} and integrating over the space. When more than $2(n-4)$ degenerations are possible the space of metrics is more complicated. There are more than $2(n-4)$ internal edges that can be chosen as a basis for the space. Making a particular choice leaves the rest as functions of the basis and requiring them to be positive gives the space more structure. 

\begin{figure}[t]
    \centering
    \includegraphics[width=\linewidth]{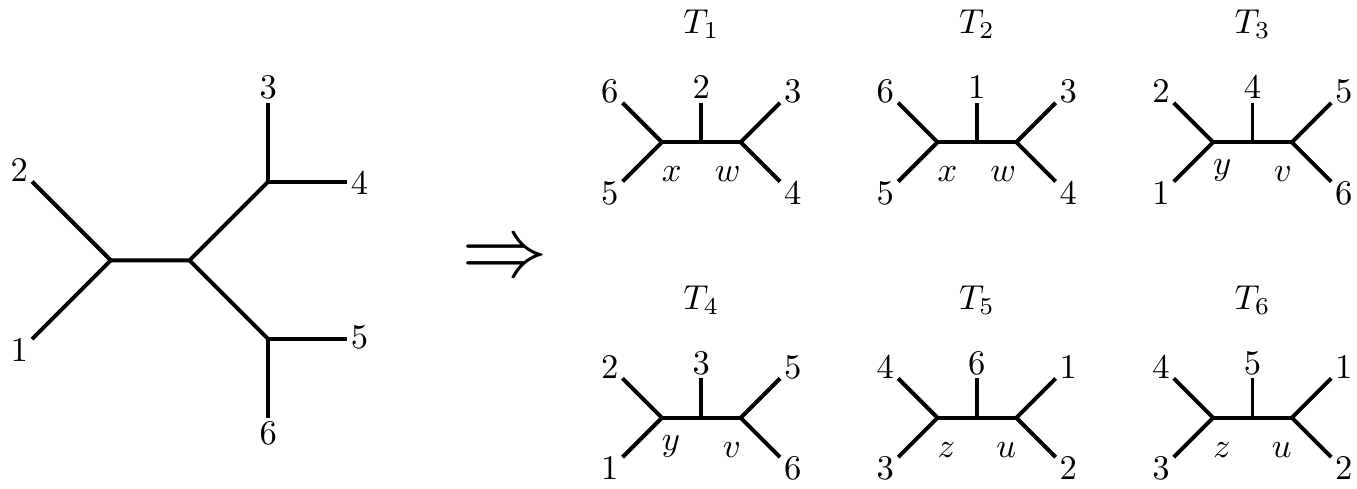}
    \caption{Snowflake tree (Left) and associated planar collection (Right).}
    \label{fig:snowflake}
\end{figure}

Let us illustrate this with an example. The simplest case where the new phenomenon, i.e. as compared to standard Feynman diagrams, arises is the $n=6$ planar collection obtained by deleting leaves from the $n=6$ snowflake tree diagram as shown in figure \ref{fig:snowflake}. We have solved the compatibility conditions on the metrics $d^{(i)}_{ab}$. Since each tree in the collection is an $n=5$ caterpillar, we can again denote the length of the first internal edge by $f^{(i)}_1$ and the second by $f^{(i)}_2$ with $i\in \{1,2,\ldots ,6\}$. In the figure we have used the solution in terms of six convenient parameters. The $2\times 6$ matrix of internal edges $f^{i}_\alpha$ is then
\be
\mqty(x & x & y & y & z & z \\ w & w & v & v & u & u)
\label{eq:snowmatrix}
\ee
subject to the constraints that $x+w=v+z$ and $y+w=u+z$. One can see that any one of the six variables can be set to zero while keeping the rest positive. These are the six possible degeneration of ${\cal T}_{\rm sf}$, where the subscript stands for snowflake.

In order to write the function ${\cal F}({\cal T}_{\rm sf})$ one has to select $2(n-4)$, i.e., four independent parameters. One possible choice is $\{x,y,z,w\}$. 
After some simple algebra one finds
\be\label{snow}
{\cal F}({\cal T}_{\rm sf}) = -\left( \sft_{1234}\,x +\sft_{3456}\, y+ (\sft_{5612}-\sfR_{654321})\, z + \sfR_{654321}\, w\right).
\ee
The kinematic invariants $\sft_{abcd}=\sfs_{abc}+\sfs_{abd}+\sfs_{acd}+\sfs_{bcd}$ and $\sfR_{654321}:=\sft_{3456}+\sfs_{341}+\sfs_{342}$ were defined in \cite{Cachazo:2019ngv} as part of the computation of generalized biadjoint amplitudes. 

The reader familiar with generalized biadjoint amplitudes would have noticed that out of the four  coefficients in \eqref{snow} only three are valid poles in an amplitude. The resolution to this naive puzzle is that the contribution of ${\cal T}_{\rm sf}$ to an amplitude is obtained by computing an integral over the space of metrics which in this case is not just $\{ x,y,z,w\}\in (\mathbb{R}^+)^4$ but the subspace constrained by the conditions that $v\geq 0$ and $u\geq 0$. More explicitly, the planar collection of Feynman diagrams' contribution to an amplitude is computed by
\be\label{formi}
{\cal R}({\cal T}_{sf})=\int_{(\mathbb{R}^+)^4} \dd{x}\,\dd{y}\,\dd{z}\,\dd{w}\,\theta(x+w-z)\theta(y+w-z)\, {\rm exp}\,{\cal F}({\cal T}_{\rm sf}), 
\ee
with $\theta(x)$ the Heaviside step function. 

These kind of integrals are all trivial if the region of integration is separated appropriately so that the step functions are removed. For example,
\be
{\cal R}({\cal T}_{sf}) =\int_{(\mathbb{R}^+)^2}\dd{y}\,\dd{w}\left(\int_{0}^y\dd{x}\int_{0}^{w+x}\dd{z} + \int_{y}^\infty\dd{x}\int_0^{w+y}\dd{z}\right)\, {\rm exp}\,{\cal F}({\cal T}_{\rm sf}),
\ee
which evaluates to
\be
{\cal R}({\cal T}_{sf}) = \frac{\sfR_{123456}+\sfR_{654321}}{\sft_{1234}\,\sft_{3456}\,\sft_{5612}\sfR_{123456}\,\sfR_{654321}}.
\ee
Note that the final formula can be written as the sum of two terms, each with four poles, or as a sum of three terms, each with four poles, by using the identity 
$\sfR_{123456}+\sfR_{654321}= \sft_{1234}+\sft_{3456}+\sft_{5612}$. This identity was noticed in \cite{Cachazo:2019ngv} and has a tropical Grassmannian version \cite{speyer2004tropical}. As an aside comment, note that this formula for ${\cal R}({\cal T}_{sf})$ exactly matches the formulation obtained in \cite{Cachazo:2019ngv} for the generalized biadjoint amplitude $m_6^{(3)}(123456,214365)$ whose only contribution in $k=2$ language comes from the snowflake Feynman diagram.

This example reveals the general structure which we express in the following definition.

\begin{definition}
The contribution of a given planar collection of Feynman diagrams, ${\cal T}$, to $m_n^{(3)}(\mathbb{I},\mathbb{I})$ is given by the integral over the space of valid metrics for ${\cal T}$ weighted by ${\rm exp}{\cal F}(\cal T)$ and is denoted by ${\cal R}({\cal T})$. More explicitly, letting $\{f_I\}$ be the set of all distinct edge lengths that define degenerations of ${\cal T}$, one can choose a subset of $2(n-4)$ independent ones, say the first ones, express the rest as functions of them and write
\be
{\cal F}({\cal T}) = \sum_{abc}\pi_{abc}\, \sfs_{abc} = -\!\!\!\sum_{I=1}^{2(n-4)}\!\! t_I f_I.
\ee
Finally, the contribution to a generalized amplitude is 
\be
{\cal R}({\cal T}):= \int_{(\mathbb{R}^+)^{2(n-4)}} \prod_{I=1}^{2(n-4)}\dd{f_I} \prod_{I=2(n-4)+1}^{{\cal D}({\cal T})}\theta \left(f_I(f_1,\ldots, f_{2(n-4)})\right)\, {\rm exp}\, {\cal F}({\cal T}).
\ee
Here ${\cal D}({\cal T})=|\{f_I\}|\geq 2(n-4)$ is the number of degenerations of the planar collection ${\cal T}$. 

\end{definition}

Even though our main interest is in  $m_n^{(3)}(\mathbb{I},\mathbb{I})$, extending the construction to $m_n^{(3)}(\alpha,\beta)$ is straightforward as it is familiar in the standard $k=2$ case. Clearly the notion of planarity with respect to $\mathbb{I}$ can be extended to any other ordering $\alpha$ and therefore
\be\label{qori}
m_n^{(3)}(\alpha,\beta) = \sum_{{\cal T}\in \Omega(\alpha) \cap \Omega(\beta)} {\cal R}({\cal T}),
\ee
with $\Omega(\alpha)$ the set of all collections of Feynman diagrams which are planar with respect to the $\alpha$-ordering.

\section{Examples}
\label{sec:examples}

In this section we illustrate the various constructions and definitions with $n=6$ and $n=7$ examples.

\subsection{Six-Point Planar Collections and Amplitudes}

The case with six points is the simplest non-trivial example. All generalized biadjoint scalar amplitudes were computed in  \cite{Cachazo:2019ngv} and reproduced from a cluster algebra point of view in \cite{Drummond:2019qjk}. Here we recompute it using planar collections of Feynman diagrams. Given that this is the most well-studied case it is important to start by mentioning the most important difference between our approach and others in the literature. In our construction we find $48$ planar collections of Feynman diagrams; $46$ of them have $4$ degenerations and two have $6$ degenerations. The contributions to $m_n^{(3)}(\mathbb{I},\mathbb{I})$ are directly computed as in the previous examples. In the literature, the way to compute the contributions is by defining an operation on simplices which are facets of ${\rm Trop}\, G(3,6)$. This is true for $46$ of them while the remaining two are bi-pyramids which are then split in either two or three simplices. Planar collections do not have to be split. 

\begin{figure}[b]
    \centering
    \includegraphics[width=\linewidth]{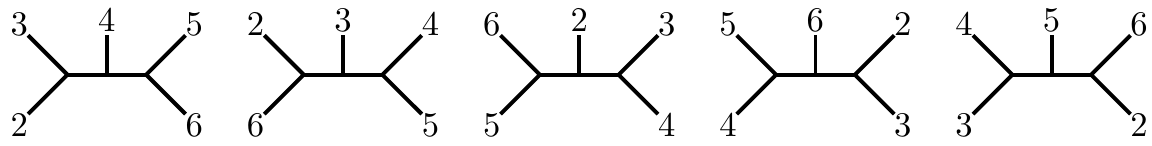}
    \caption{Five possible planar graph for the first tree of any six-point planar collection.}
    \label{fig:firstree}
\end{figure}

Let us start the study of $n=6$ planar collections of Feynman diagrams by introducing a very convenient notation. In this case each collection is a set of six trees with five leaves. Each tree in the family is chosen from one of the five possible planar graphs. For example, the first tree in any collection is one of the five trees shown in figure \ref{fig:firstree}. Note that planarity again simplifies the problem and that each tree is uniquely labelled by the leaf in the middle, i.e., the leaf that does not belong to any of the four cherries. This observation motivates the following notation.

A $n=6$ planar collection ${\cal T} := [a_1,a_2,a_3,a_4,a_5,a_6]$ has as the $i^{\rm th}$ tree the planar caterpillar $C(a_i-2,a_i-1,a_i,a_i+1,a_i+2)$ where indices are taking cyclically in the order $\{1,2,3,4,5,6\}\setminus i$.

Using this notation the list of all $48$ planar collection is presented in table \ref{tb:n6coll}. In order to check our constructions we first computed the $48$ planar collections by listing all $5^6=15\, 625$ possible combinations $\{ a_1,a_2,a_3,a_4,a_5,a_6\}$ and then check which ones admit metrics satisfying the compatibility condition. Bringing the number down to $48$ shows how restrictive the condition is. We review an efficient way of doing this for $n<9$ using abstract tree arrangements in appendix \ref{ap:treearrs}.

\begin{table}[t]
\centering
\begin{tabular}{|c|c||c|c|}
\hline 
\multicolumn{4}{|c|}{Planar collections of trees in $k=3$ and $n=6$} \\ 
\hline 
Collection & Trees & Collection & Trees \\ 
\hline 
$\mathcal{T}_{1}$ & $\qty[4,4,4,3,3,3]$ & $\mathcal{T}_{25}$ & $\qty[6,6,6,5,4,1]$ \\ 
\hline 
$\mathcal{T}_{2}$ & $\qty[4,4,4,3,6,5]$ & $\mathcal{T}_{26}$ & $\qty[6,6,6,6,6,3]$ \\ 
\hline 
$\mathcal{T}_{3}$ & $\qty[4,4,4,3,2,2]$ & $\mathcal{T}_{27}$ & $\qty[6,6,6,1,1,1]$ \\ 
\hline 
$\mathcal{T}_{4}$ & $\qty[4,4,4,1,4,4]$ & $\mathcal{T}_{28}$ & $\qty[6,6,6,2,2,1]$ \\ 
\hline 
$\mathcal{T}_{5}$ & $\qty[4,4,4,1,1,1]$ & $\mathcal{T}_{29}$ & $\qty[6,3,2,5,4,1]$ \\ 
\hline 
$\mathcal{T}_{6}$ & $\qty[4,4,6,6,6,5]$ & $\mathcal{T}_{30}$ & $\qty[6,3,2,1,1,1]$ \\ 
\hline 
$\mathcal{T}_{7}$ & $\qty[4,4,6,6,2,2]$ & $\mathcal{T}_{31}$ & $\qty[6,3,2,2,2,1]$ \\ 
\hline 
$\mathcal{T}_{8}$ & $\qty[4,5,5,5,4,4]$ & $\mathcal{T}_{32}$ & $\qty[2,5,5,5,2,2]$ \\ 
\hline 
$\mathcal{T}_{9}$ & $\qty[4,6,6,5,4,4]$ & $\mathcal{T}_{33}$ & $\qty[2,5,2,2,2,2]$ \\ 
\hline 
$\mathcal{T}_{10}$ & $\qty[4,6,6,2,2,4]$ & $\mathcal{T}_{34}$ & $\qty[2,1,4,3,3,3]$ \\ 
\hline 
$\mathcal{T}_{11}$ & $\qty[4,1,1,1,4,4]$ & $\mathcal{T}_{35}$ & $\qty[2,1,4,3,6,5]$ \\ 
\hline 
$\mathcal{T}_{12}$ & $\qty[4,1,1,1,1,1]$ & $\mathcal{T}_{36}$ & $\qty[2,1,4,3,2,2]$ \\ 
\hline 
$\mathcal{T}_{13}$ & $\qty[4,3,2,5,4,4]$ & $\mathcal{T}_{37}$ & $\qty[2,1,6,6,6,5]$ \\ 
\hline 
$\mathcal{T}_{14}$ & $\qty[4,3,2,2,2,4]$ & $\mathcal{T}_{38}$ & $\qty[2,1,6,6,2,2]$ \\ 
\hline 
$\mathcal{T}_{15}$ & $\qty[5,5,4,3,3,3]$ & $\mathcal{T}_{39}$ & $\qty[2,1,1,1,3,3]$ \\ 
\hline 
$\mathcal{T}_{16}$ & $\qty[5,5,4,3,6,5]$ & $\mathcal{T}_{40}$ & $\qty[2,1,1,1,6,5]$ \\ 
\hline 
$\mathcal{T}_{17}$ & $\qty[5,5,5,5,2,5]$ & $\mathcal{T}_{41}$ & $\qty[2,1,1,1,2,2]$ \\ 
\hline 
$\mathcal{T}_{18}$ & $\qty[5,5,6,6,6,5]$ & $\mathcal{T}_{42}$ & $\qty[3,5,5,5,4,3]$ \\ 
\hline 
$\mathcal{T}_{19}$ & $\qty[5,5,1,1,3,3]$ & $\mathcal{T}_{43}$ & $\qty[3,5,5,1,1,3]$ \\ 
\hline 
$\mathcal{T}_{20}$ & $\qty[5,5,1,1,6,5]$ & $\mathcal{T}_{44}$ & $\qty[3,3,6,3,3,3]$ \\ 
\hline 
$\mathcal{T}_{21}$ & $\qty[5,5,2,2,2,5]$ & $\mathcal{T}_{45}$ & $\qty[3,3,6,6,6,3]$ \\ 
\hline 
$\mathcal{T}_{22}$ & $\qty[6,5,5,5,4,1]$ & $\mathcal{T}_{46}$ & $\qty[3,3,2,5,4,3]$ \\ 
\hline 
$\mathcal{T}_{23}$ & $\qty[6,5,5,1,1,1]$ & $\mathcal{T}_{47}$ & $\qty[3,3,2,1,1,3]$ \\ 
\hline 
$\mathcal{T}_{24}$ & $\qty[6,6,6,3,3,3]$ & $\mathcal{T}_{48}$ & $\qty[3,3,2,2,2,3]$ \\ 
\hline 
\end{tabular}
\caption{All $48$ planar collections of trees for $n=6$ in a compact notation tailored to this case and explained in the text.}
\label{tb:n6coll}
\end{table}

\begin{figure}[t]
    \centering
    \includegraphics[width=0.95\linewidth]{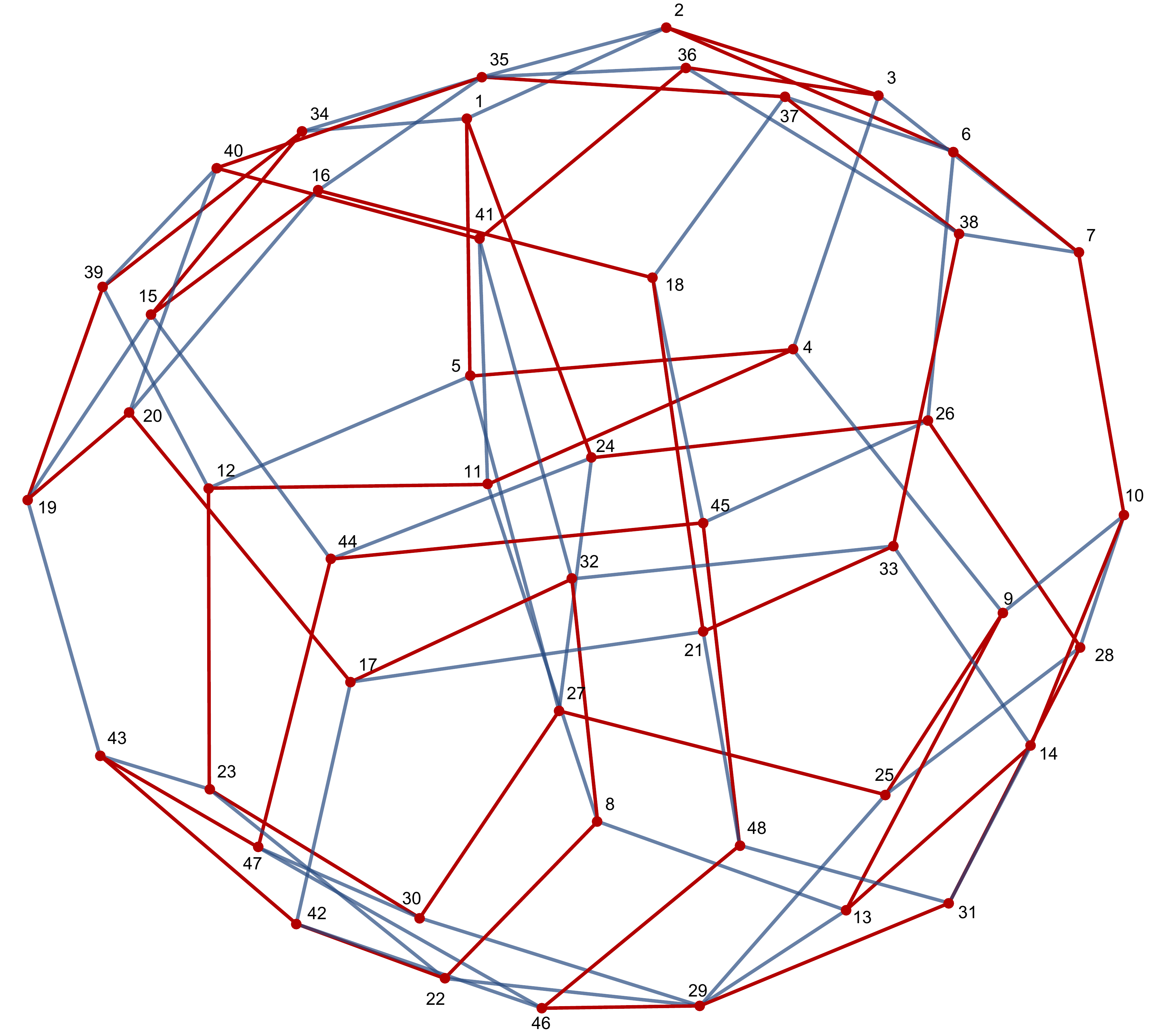}
    \caption{Each vertex represents a planar collection. The degree of a vertex is the number of degenerations. There are $46$ vertices of degree $4$ and $2$ vertices, i.e. $v_{29}$ and $v_{35}$, of degree six. The Hamiltonian cycle used in this work is highlighted in red.}
    \label{fig:hamcycle}
\end{figure}

The next step is to check that indeed all $48$ planar collections can be generated from a single one, e.g. ${\cal T}_1$ by traveling through degenerations. Since $n=6$ is still small enough for complete computations, we followed all degenerations, four for all collections except for ${\cal T}_{29}$ and ${\cal T}_{35}$ which have six, and constructed the graph of connections in figure \ref{fig:hamcycle}. Using the graph it is easy to find a Hamiltonian cycle.

The Hamiltonian cycle we use is:
\begin{equation}
\begin{aligned}
{\cal T}_1 &\to {\cal T}_5 \to {\cal T}_4 \to {\cal T}_{11} \to {\cal T}_{12} \to {\cal T}_{23} \to {\cal T}_{30} \to {\cal T}_{27} \to {\cal T}_{25} \to {\cal T}_9 \to {\cal T}_{13} \to {\cal T}_{14} \to {\cal T}_{10} \\
&\to {\cal T}_7 \to {\cal T}_6 \to {\cal T}_2 \to {\cal T}_3 \to {\cal T}_{36} \to {\cal T}_{41} \to {\cal T}_{40} \to {\cal T}_{35} \to {\cal T}_{37} \to {\cal T}_{38} \to {\cal T}_{33} \to {\cal T}_{21} \\
&\to {\cal T}_{18} \to {\cal T}_{16} \to {\cal T}_{15} \to {\cal T}_{34} \to {\cal T}_{39} \to {\cal T}_{19} \to {\cal T}_{20} \to {\cal T}_{17} \to {\cal T}_{32} \to {\cal T}_8 \to {\cal T}_{22} \to {\cal T}_{42} \\
&\to {\cal T}_{43} \to {\cal T}_{43} \to {\cal T}_{47} \to {\cal T}_{44} \to {\cal T}_{45} \to {\cal T}_{48} \to {\cal T}_{46} \to {\cal T}_{29} \to {\cal T}_{31} \to {\cal T}_{28} \to {\cal T}_{26}
\to {\cal T}_{24} \\
&\to {\cal T}_1
\end{aligned}
\end{equation}

It turns out that the $48$ planar collections separate even further into six classes depending on the kind of poles their contribution to $m_6^{(3)}(\mathbb{I},\mathbb{I})$ has. Therefore, in order to recompute $m_6^{(3)}(\mathbb{I},\mathbb{I})$ using our approach we present the matrices of internal edges for a representative of all six distinct collections and their corresponding contribution to the amplitude: 

\begin{itemize}
	\item $\sfs\,\sfs\,\sfs\,\sfR$: Example $\mathcal{T}_{10}=\qty[4,6,6,2,2,4]$

	\begin{subequations}
	Matrix and constraints
	\begin{align}
	\mqty(x & y & z & y & w & y \\ v & u & u & p & p & v)\qc & \mqty{y + u = x, \\ y + p = z, \\ y + v = w.}
	\end{align}
	The function $\mathcal{F}(\mathcal{T}_{10})$ and contribution to the amplitude $\mathcal{R}(\mathcal{T}_{10})$ are
	\begin{gather}
	    \mathcal{F}(\mathcal{T}_{10}) = -\qty(\sfs_{123}\,u + \sfs_{345}\,p + \sfs_{561}\,v + \sfR_{165432}\,y),\\
	    \mathcal{R}(\mathcal{T}_{10}) = \frac{1}{\sfs_{123}\sfs_{345}\sfs_{561}\sfR_{165432}}.
	\end{gather}
	Here and below we ignore irrelevant numerical overall constants in $\mathcal{F}(\mathcal{T})$.
	\end{subequations}
	\item $\sfs\,\sfs\,\sft\,\sft(a)$: Example $\mathcal{T}_{1}=\qty[4,4,4,3,3,3]$
	
	\begin{subequations}
	Matrix and constraints
	\begin{equation}
	\mqty(x & x & y & z & z & z \\ w & w & w & v & u & u )\qc \mqty{x + z = y, \\ w + u = v.}
	\end{equation}
	The function $\mathcal{F}(\mathcal{T}_{1})$ and contribution to the amplitude $\mathcal{R}(\mathcal{T}_{1})$ are
	\begin{gather}
    \mathcal{F}(\mathcal{T}_{1}) = -\qty(\sfs_{123}\,x + \sfs_{456}\,u + \sft_{1234}\,w + \sft_{3456}\,z),\\
	    \mathcal{R}(\mathcal{T}_{1}) = \frac{1}{\sfs_{123}\sfs_{456}\sft_{1234}\sft_{3456}}.
	\end{gather}
	\end{subequations}
	\item $\sfs\,\sfs\,\sft\,\sft(b)$: Example $\mathcal{T}_{4}=\qty[4,4,4,1,4,4]$
	
	\begin{subequations}
	Matrix and constraints
	\begin{equation}
	\mqty(x & y & y & z & w & w \\ v & z & z & w & u & u)\qc \mqty{y + w = x, \\ z + u = v.}
	\end{equation}
	The function $\mathcal{F}(\mathcal{T}_{4})$ and contribution to the amplitude $\mathcal{R}(\mathcal{T}_{4})$ are
	\begin{gather}
	    \mathcal{F}(\mathcal{T}_{4}) = -\qty(\sfs_{123}\,y + \sfs_{561}\,u + \sft_{1234}\,z + \sft_{4561}\,w),\\
	    \mathcal{R}(\mathcal{T}_{4}) = \frac{1}{\sfs_{123}\sfs_{561}\sft_{1234}\sft_{4561}}.
	\end{gather}
	\end{subequations}
	
	\item $\sfs\,\sfs\,\sft\,\sfR$: Example $\mathcal{T}_{23}=\qty[6,5,5,1,1,1]$
	
	\begin{subequations}
	Matrix and constraints
	\begin{equation}
	\mqty(x & y & y & z & z & w \\ v & x & x & u & p & v)\qc \mqty{x + v = p,\\ y + p = u, \\ x + z = w.}
	\end{equation}
	The function $\mathcal{F}(\mathcal{T}_{23})$ and contribution to the amplitude $\mathcal{R}(\mathcal{T}_{23})$ are
	\begin{gather}
	    \mathcal{F}(\mathcal{T}_{23}) = -\qty(\sfs_{234}\,y + \sfs_{456}\,z + \sft_{4561}\,v + \sfR_{456123}\,x),\\
	    \mathcal{R}(\mathcal{T}_{23}) = \frac{1}{\sfs_{234}\sfs_{456}\sft_{4561}\sfR_{456123}}.
	\end{gather}
	\end{subequations}

	\item $\sfs\,\sft\,\sft\,\sfR$ : Example $\mathcal{T}_{30}=\qty[6,3,2,1,1,1]$
	
	\begin{subequations}
	Matrix and constraints
	\begin{equation}
	\mqty(x & y & y & z & z & w \\ v & u & u & p & p & v)\qc \mqty{x + z = w, \\ y + v = p, \\ y + u = x.}
	\end{equation}
	The function $\mathcal{F}(\mathcal{T}_{30})$ and contribution to the amplitude $\mathcal{R}(\mathcal{T}_{30})$ are
	\begin{gather}
	    \mathcal{F}(\mathcal{T}_{30}) = -\qty(\sfs_{456}\,z + \sft_{4561}\,v + \sft_{6123}\,u + \sfR_{612345}\,y),\\
	    \mathcal{R}(\mathcal{T}_{30}) = \frac{1}{\sfs_{456}\sft_{4561}\sft_{6123}\sfR_{612345}}.
	\end{gather}
	\end{subequations}
	\item $\sft\,\sft\,\sft\,\sfR\,\sfR$ : Example $\mathcal{T}_{35}=\qty[2,1,4,3,6,5]$
	
	\begin{subequations}
	Matrix and constraints
	\begin{equation}
	\mqty(x & x & y & y & z & z \\ w & w & v & v & u & u)\qc \mqty{x+w=v+z, \\ y+w=u+z.}
	\end{equation}
	The function $\mathcal{F}(\mathcal{T}_{35})$ and contribution to the amplitude $\mathcal{R}(\mathcal{T}_{35})$ are
	\begin{gather}
	    \mathcal{F}(\mathcal{T}_{35}) = -\qty( \sft_{1234}\,x +\sft_{3456}\, y+ (\sft_{5612}-\sfR_{654321})\, z+ \sfR_{654321}\, w),\\
	    \mathcal{R}(\mathcal{T}_{35}) = \frac{\sft_{1234}+\sft_{3456}+\sft_{5612}}{\sft_{1234}\sft_{3456}\sft_{5612}\sfR_{123456}\sfR_{654321}}.
	\end{gather}
	This example is the bi-pyramid computed at the beginning of section \ref{sec:genamps} (see eqs. \eqref{eq:snowmatrix}-\eqref{formi}). Notice that this example is the only one in the list with a non-trivial $\mathcal{F}$ function (only three of the coefficients are valid poles in the amplitude).
	\end{subequations}
\end{itemize}


\subsection{Seven-Point Planar Collections}
\label{sec:sevenpexample}

Seven-point planar collections are much richer than those for $n=6$ and also more numerous. We have also found all $693$ planar collections. These can be classified by the number of poles in ${\cal R}({\cal T})$ into four classes. These contain $6,7,8$ and $9$ poles and there are $595$, $63$, $28$ and $7$ planar collections in each class respectively. In section \ref{sec:clusterconnection} we discuss the connection to cluster algebras and ${\rm Trop}^+\, G(3,7)$ where the same decomposition of facets happens with poles counted as vertices.

In $n=7$ there are five different kind of poles. These are $\sfs_{abc}$ and $\sft_{abcd}$ with the same definitions as for $n=6$, and the remaining three are \cite{Cachazo:2019apa}:
\be
\begin{gathered}
\sft_{abcde}= \sfs_{abc}+\sfs_{abd}+\sfs_{abe}+\sfs_{acd}+\sfs_{ace}+\sfs_{ade}+\sfs_{bcd}+\sfs_{bce}+\sfs_{bde}+\sfs_{cde},\\
\sfR_{abcdefg} = \sft_{abcde}+\sfs_{def}+\sfs_{deg}\qc \sfW_{abcdefg} = \sft_{abcd} + \sft_{fgab} + \sfs_{abe}.
\end{gathered}
\ee
Here we have changed the definition of $\sfW$ with respect to \cite{Cachazo:2019apa} to make planarity explicit\footnote{The original definition was $\sfW_{abcdefg} = \sft_{abcd} + \sft_{abef} + \sfs_{abg}$. A planar example in this definition is $\sfW_{1234675}$.}.

\begin{figure}[t]
    \centering
    \includegraphics[width=\linewidth]{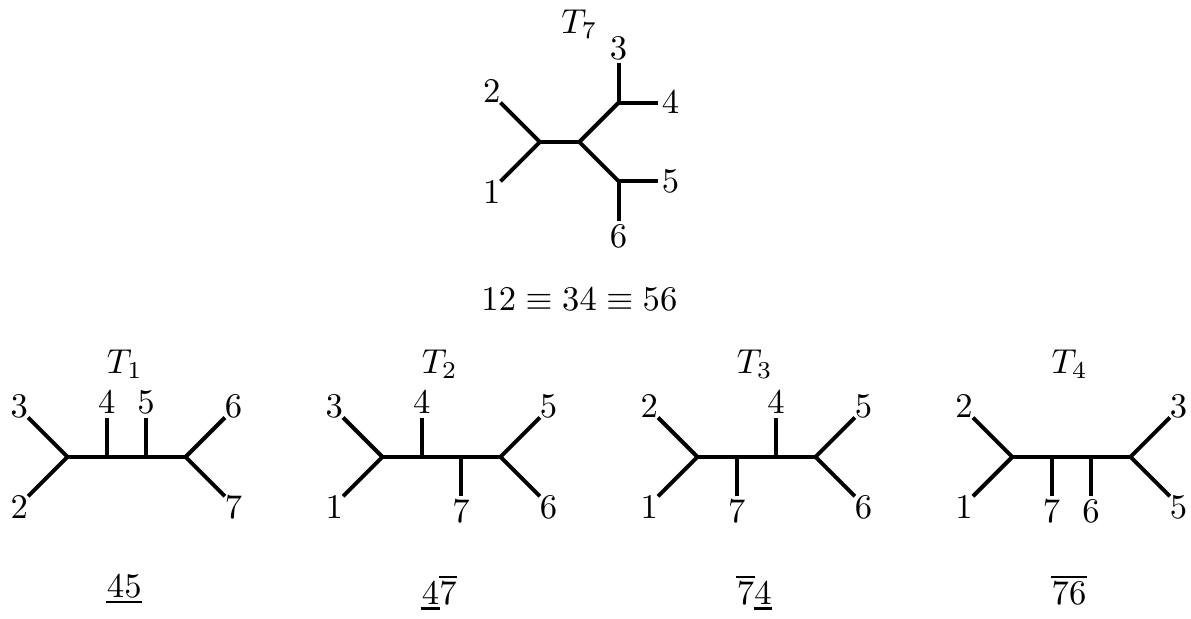}
    \caption{Notation used to identify planar trees in $k=3$, $n=7$. Top: Planar snowflakes can be identified by any of their cherries. Bottom: Planar Caterpillar trees can be identified by their two internal lengths and how they are planarly related to the cherries. Again, the position of any tree (given by $T_a$) in the collection determines the label not present in the tree.}
    \label{fig:seventopol}
\end{figure}
Let us introduce a compact notation for $n=7$ planar collections. Each collection is made out of six-point planar trees which come in two basic topologies: caterpillars and snowflakes. Planar snowflakes are completely determined by the two labels on any of its three cherries. Caterpillars come in several varieties depending on whether leaves are on one side or the other of the main body (this is only relevant due to planarity). The example in figure \ref{fig:seventopol} shows the labeling for both topologies. The snowflake has $1$ and $2$ in the same cherry and can denoted as $12$; likewise, it can also be denoted $34$ or $56$. In the same figure, we can see all possible labeling for caterpillars. In the tree $T_2$ for example, $4$ is on top and $7$ at the bottom, therefore we use the notation $\underline{4}\overline{7}$.

We do not provide a full list of planar collections here. Instead we give a sample in table \ref{tb:sevenpointfam} which also contains examples of each class. In table \ref{tb:sevenpointfam}, each class contributes with 7 collections that can be spanned by taking successive cyclic translations on the labels and positions of each tree in the collection. Let $(ab,cd,ef)$ denote the leaves and legs of the $g\textsuperscript{th}$ tree in a given collection $\mathcal{T}$ ($T_g$ can be a caterpillar, e.g. $\overline{c}\underline{d}$ or snowflake, e.g. $ab$), the translation operation described in the last sentence is equivalent to:
\be
\begin{aligned}
T_g &\mapsto T_{(g\,\mathrm{mod}\,7)+1},\\
(ab,cd,ef) &\mapsto \qty((a\,\mathrm{mod}\,7)+1, (b\,\mathrm{mod}\,7)+1, \ldots.)
\end{aligned}
\ee

\begin{table}[t]
\centering
\begin{tabular}{|c|c|c|}
\hline 
\multicolumn{3}{|c|}{Planar collections of trees in $k=3$ and $n=7$} \\ 
\hline 
N. of Poles & Class (Representative) & Collection \\ 
\hline 
\multirow{6}{*}{6} & $\mathcal{T}_{1}$ & $[\overline{7}\underline{4},\overline{7}\underline{4},\overline{76},\underline{2}\overline{6},\underline{12},\underline{12},\overline{1}\underline{4}]$ \\ 
\cline{2-3} 
 & $\mathcal{T}_{2}$ & $[\overline{65},\overline{65},\overline{5}\underline{1},\underline{71},\underline{7}\overline{3},\underline{7}\overline{3},\underline{6}\overline{3}]$ \\ 
\cline{2-3}
 & \multicolumn{2}{c|}{\vdots} \\ 
\cline{2-3} 
 & $\mathcal{T}_{84}$ & $[\underline{5}\overline{2},\underline{5}\overline{1},\underline{45},\underline{35},12,12,12]$ \\ 
\cline{2-3}
 & $\mathcal{T}_{85}$ & $[\overline{27},\overline{17},\overline{7}\underline{4},\overline{7}\underline{3},12,12,12]$ \\ 
\hline 
\multirow{5}{*}{7} & $\mathcal{T}_{86}$ & $[\overline{2}\underline{5},\overline{1}\underline{5},\underline{45},\underline{3}\overline{7},\underline{3}\overline{7},\overline{7}\underline{3},\overline{6}\underline{3}]$ \\ 
\cline{2-3}
 & $\mathcal{T}_{87}$ & $[\overline{2}\underline{5},\overline{1}\underline{5},\underline{4}\overline{7},\underline{3}\overline{7},\overline{7}\underline{3},\overline{7}\underline{3},\overline{65}]$ \\ 
\cline{2-3} 
 & \multicolumn{2}{c|}{\vdots} \\ 
\cline{2-3}
 & $\mathcal{T}_{93}$ & $[\overline{2}\underline{5},\overline{1}\underline{5},\underline{45},\underline{35},12,12,12]$ \\ 
\cline{2-3} 
 & $\mathcal{T}_{94}$ & $[\overline{27},\overline{17},\underline{4}\overline{7},\underline{3}\overline{7},12,12,12]$ \\ 
\hline 
\multirow{4}{*}{8} & $\mathcal{T}_{95}$ & $[\overline{2}\underline{5},\overline{1}\underline{5},\underline{4}\overline{7},\underline{3}\overline{7},\overline{7}\underline{3},\overline{7}\underline{3},\overline{65}]$ \\ 
\cline{2-3} 
 & $\mathcal{T}_{96}$ & $[\overline{2}\underline{5},\overline{1}\underline{5},\underline{4}\overline{7},\underline{3}\overline{7},\underline{67},\underline{5}\overline{2},\underline{5}\overline{2}]$ \\ 
\cline{2-3}
 & $\mathcal{T}_{97}$ & $[\overline{2}\underline{5},\overline{1}\underline{5},\underline{4}\underline{5},\underline{3}\underline{5},12,\underline{5}\underline{7},\underline{5}\underline{6}]$ \\ 
\cline{2-3}
 & $\mathcal{T}_{98}$ & $[\underline{23},\underline{13},12,\overline{53},\overline{43},\underline{7}\overline{3},\underline{6}\overline{3}]$ \\ 
\hline 
9 & $\mathcal{T}_{99}$ & $[\overline{2}\underline{5},\overline{1}\underline{5},\underline{4}\overline{7},\underline{3}\overline{7},\overline{76},\overline{75},\overline{65}]$ \\ 
\hline 
\end{tabular} 
\caption{Some representative collections from the contributing planar families in the $n=7$ planar amplitude. Each representative belongs to a defined family of planar arrangements, each of which contribute with a number of planar collections shown in the rightmost column.}
\label{tb:sevenpointfam}
\end{table}


Here we concentrate on illustrating how the contribution to an amplitude is computed for one example from each of the five classes distinguished by the number of poles in the answer.

The first class is that with six poles. This is the generic case since it corresponds to planar collections with exactly $2(n-4)=6$ degenerations. The planar collection caterpillar collection presented in section \ref{sec:allmult} for all $n$ is an example.   

\subsubsection{Planar Collection ${\cal T}_{86}$: Seven Poles}

The planar collection, in the compact notation introduced above, is given by
\be
{\cal T}_{86} =[\overline{2}\underline{5}, \overline{1}\underline{5}, \underline{45}, \underline{3}\overline{7}, \underline{3}\overline{7}, \overline{7}\underline{3}, \overline{6}\underline{3}].
\ee

The matrix of internal lengths is
\be
\left(
\begin{array}{ccccccc}
 x & x & y & y & x+y & z & z \\
 w & w & w+x & p+w+x & p & q & q \\
 u & u & p & v & v & p+v & p \\
\end{array}
\right),
\ee
with 
\be
p= u - x - y + z, \quad q= x + y - z.
\ee
This means that this planar collection has eight degenerations. Of course, there are only six parameters, $\{x,y,z,w,u,v\}$ as expected since $2(n-4)=6$ for $n=7$.

Let us compute the contribution to the amplitude. The first step is the function
\be
{\cal F}({\cal T}_{86}) = -\left( k_x\, x+k_y\, y+ k_z\, z+k_w\, w+k_v\, v+k_u\, u \right)
\ee
with 
\be\begin{gathered}
k_x:=\sfW_{1234567}-\sfR_{1234567}\qc k_y:=\sfR_{3456712}-\sfR_{1234567}\qc k_z:=\sfR_{7654321}-k_x,\\ k_w:=\sft_{1234} \qc k_v:=\sfR_{1234567}\qc k_u:=\sfs_{456}.
\end{gathered}
\ee
The next step is to compute the integral over the space of metrics, recalling that one must impose $p\geq 0$ and $q\geq 0$,
\be
{\cal R}({\cal T}_{86}) =\int_{(\mathbb{R}^+)^6}\dd{x}\,\dd{y}\,\dd{z}\,\dd{w}\,\dd{v}\,\dd{u}\,\theta(u - x - y + z)\theta(x + y - z)\, {\rm exp}\,{\cal F}({\cal T}_{86}).
\ee
The integral can be easily done by using the function $\texttt{Reduce}$ in $\texttt{Mathematica}$ on all the inequalities defining the space,
\be
x\geq0,\; y\geq 0,\; \ldots ,u\geq 0,\; u - x - y + z\geq 0,\; x+y\geq z
\ee
to remove the step functions and replace them by new regions of integration:
\be
{\cal R}({\cal T}_{86}) =\int_{(\mathbb{R}^+)^4}\dd{x}\,\dd{y}\,\dd{w}\,\dd{v}\int_{0}^{x+y}\dd{z}\int_{x+y-z}^{\infty}\dd{u}\, {\rm exp}\,{\cal F}({\cal T}_{86}).
\ee
The result of this integral is the final result
\be
{\cal R}({\cal T}_{86}) = \frac{\sfW_{1234567}+\sft_{34567}}{\sfs_{456}\sft_{1234}\sft_{34567}\sfR_{1234567}\sfR_{3456712}\sfR_{7654321}\sfW_{1234567}}.
\ee

\subsubsection{Planar Collection ${\cal T}_{97}$: Eight Poles}

The planar collection is given by
\be
{\cal T}_{97} = [\overline{2}\underline{5},\overline{1}\underline{5},\underline{4}\underline{5},\underline{3}\underline{5},12,\underline{5}\underline{7},\underline{5}\underline{6}].
\ee

The matrix of internal lengths is
\be
\left(
  \begin{array}{ccccccc}
    x & x & y & y & z+u & z & z \\
    w & w & p & p & q & u & u \\
    v+u & v+u & v & v & v & q & q \\
  \end{array}
\right),
\ee
with
\be
p=z+w+u-y,\qquad q = x+y-z-u.
\ee
This means that this planar collection has eight degenerations. 

Let us compute the contribution to the amplitude. The first step is the function
\be
{\cal F}({\cal T}_{97}) = -\left( k_x\, x+k_y\, y+ k_z\, z+k_w\, w+k_v\, v+k_u\, u \right)
\ee
with
\be\begin{gathered}
k_x:=\sft_{56712}\qc k_y:=\sfR_{5671234}-\sft_{1234}\qc k_z:=\sft_{34567}+\sft_{1234}-\sfR_{5671234},\\ k_w:=\sft_{1234} \qc k_v:=\sft_{12345}\qc k_u:=\sfW_{1234567}-\sft_{56712}.
\end{gathered}
\ee
The next step is to compute the integral over the space of metrics, recalling that one must impose $p\geq 0$ and $q\geq 0$,
\be
{\cal R}({\cal T}_{97}) =\int_{(\mathbb{R}^+)^6}\dd{x}\,\dd{y}\,\dd{z}\,\dd{w}\,\dd{v}\,\dd{u}\,\theta(z+w+u-y)\theta(x+y-z-u)\, {\rm exp}\,{\cal F}({\cal T}_{97}).
\ee
This integral can again be easily performed by separating into regions the integration domain using the function $\texttt{Reduce}$ in $\texttt{Mathematica}$ on all the inequalities defining the space, i.e. 
\be
x\geq0,\; y\geq 0,\; \ldots ,u\geq 0,\; z+w+u\geq y,\; x+y\geq z+u.
\ee
The integral over the first region, $0\leq z\leq y$, can be written as
\be
{\cal R}_1({\cal T}_{97}) =\! \int_{(\mathbb{R}^+)^3}\!\!\dd{x}\,\dd{y}\,\dd{v}\int_{0}^{y}\dd{z}\left(\int_{0}^{y-z}\! \dd{w}\int_{y-z-w}^{x+y-z}\! \dd{u}+\int_{y-z}^{\infty}\! \dd{w}\int_{0}^{x+y-z}\! \dd{u}\right)\, {\rm exp}\,{\cal F}({\cal T}_{97})
\ee
and it is straightforward to evaluate the two parts, i.e. ${\cal R}_1({\cal T}_{97}) ={\cal R}_1^{(1)} +{\cal R}_1^{(2)}$, with
\be
{\cal R}_1^{(1)} = \frac{\sfR_{5671234}+\sfW_{1234567}}{\sft_{12345}\sft_{34567}\sft_{56712}\sfR_{3456712}\sfR_{5671234}\sfW_{1234567}(\sfR_{5671234}+\sft_{1234})}.
\ee
\be
{\cal R}_1^{(2)} = \frac{\sfR_{5671234}+\sfW_{1234567}}{\sft_{1234}\sft_{12345}\sft_{34567}\sft_{56712}\sfR_{5671234}\sfW_{1234567}(\sfR_{5671234}+\sft_{1234})}.
\ee
Note that these two functions contain the spurious pole $\sfR_{5671234}+\sft_{1234}$ but neatly combine to cancel it into
\be
{\cal R}_1({\cal T}_{97}) = \frac{\sfR_{5671234}+\sfW_{1234567}}{\sft_{1234}\sft_{12345}\sft_{34567}\sft_{56712}\sfR_{3456712}\sfR_{5671234}\sfW_{1234567}}.
\ee

The integral over the second region, $y\leq z\leq x+y$, is even simpler,
\be
\begin{aligned}
{\cal R}_2({\cal T}_{97})  &= \int_{(\mathbb{R}^+)^4}\dd{x}\,\dd{y}\,\dd{v}\,\dd{w}\int_{y}^{x+y}\dd{z}\int_{0}^{x+y-z}\dd{u} {\rm exp}\,{\cal F}({\cal T}_{97}) \\ &=\frac{1}{\sft_{1234}\sft_{12345}\sft_{34567}\sft_{56712}\sfR_{7654321}\sfW_{1234567}}.
\end{aligned}
\ee
Finally, since ${\cal R}({\cal T}_{97})={\cal R}_1({\cal T}_{97})+{\cal R}_2({\cal T}_{97})$ one could combine the results into a single object. Instead it is interesting to note that ${\cal R}_2$ has the structure a planar collection with exactly six degenerations would have had. Moreover,  ${\cal R}_1({\cal T}_{97})$ is the sum of two such rational functions. Combining these results
\be\label{simT}
{\cal R}({\cal T}_{97}) =\! \frac{1}{\sft_{1234}\sft_{12345}\sft_{34567}\sft_{56712}}\!\left( \frac{1}{\sfR_{3456712}\sfR_{5671234}}+\frac{1}{\sfR_{3456712}\sfW_{1234567}} +\frac{1}{\sfR_{7654321}\sfW_{1234567}} \right).
\ee

As reviewed in section \ref{sec:clusterconnection}, the same object, but thought of as a facet of ${\rm Trop}^+\, G(3,7)$, is given by three clusters. It would be interesting to compare the three-term formula \eqref{simT} with the one obtained from the three clusters.

\subsubsection{Planar Collection ${\cal T}_{99}$: Nine Poles}

The planar collection, in the compact notation introduced above, is given by
\be
{\cal T}_{99} =[\overline{2}\underline{5},\underline{1}\overline{5},\underline{4}\overline{7},\underline{3}\overline{7},\overline{76},\overline{75},\overline{65} ].
\ee

The matrix of internal edges is given by
\be
\left(
\begin{array}{ccccccc}
 x & x & y & y & z & z-v & z-v \\
 w & w & w-y+z & w-y+z & u & u+v & v \\
 v & v & u & u & -u+x+y-z & -u+x+y-z & x+y-z \\
\end{array}
\right).
\ee
This means that this planar collection could have up to $10$ degenerations. However, it is easy to check that if the last entry of the third row $x+y-z\to 0$, its neighbor to the left becomes $-u$ and hence this is not a valid degeneration. Therefore, there are only $9$ valid degenerations.

Let us compute the contribution to the amplitude. The first step is the function
\be
{\cal F}({\cal T}_{99}) = -\left( k_x\, x+k_y\, y+ k_z\, z+k_w\, w+k_v\, v+k_u\, u \right)
\ee
with
\be\begin{gathered}
k_x:=\sft_{56712}\qc k_y:=-\sfR_{7654321}+\sft_{34567}+\sft_{56712}\qc k_z:=\sfR_{7654321}-\sft_{56712},\\ k_w:=\sft_{1234} \qc k_v:=-\sfR_{7654321}+\sfW_{1234567}\qc k_u:=\sfR_{2176543}-\sft_{56712}.
\end{gathered}
\ee
The next step is to compute the integral over the space of metrics,
\be
{\cal R}({\cal T}_{99}) =\int_{(\mathbb{R}^+)^6}\dd{x}\,\dd{y}\,\dd{z}\,\dd{w}\,\dd{v}\,\dd{u}\,\theta(z-v)\theta(w-y+z)\theta(-u+x+y-z)\, {\rm exp}\,{\cal F}({\cal T}_{99}).
\ee
Once again, this integral can be easily performed by separating into two regions the integration domain:
\be
{\cal R}_1({\cal T}_{99}) =\int_{(\mathbb{R})^2}\dd{x}\,\dd{y}\int_{0}^y\dd{z}\int_{y-z}^{\infty}\dd{w}\int_0^{x+y-z}\dd{u}\int_0^z \dd{v}\, {\rm exp}\,{\cal F}({\cal T}_{99}).
\ee
which gives
\be
{\cal R}_1({\cal T}_{99}) = \frac{\sfR_{5671234}+\sfR_{2176543}}{\sft_{1234}
\sft_{56712}
\sft_{34567}
\sfR_{3456712}
\sfR_{2176543}
\sfR_{5671234}
\sfW_{4321765}}.
\ee
The second domain leads to the following integral
\be
{\cal R}_2({\cal T}_{99}) =\int_{(\mathbb{R})^3}\dd{x}\,\dd{y}\,\dd{w}\int_{y}^{x+y}\dd{z}\int_0^{x+y-z}\dd{u}\int_0^z \dd{v}\, {\rm exp}\,{\cal F}({\cal T}_{99}).
\ee
which evaluates to
\be
{\cal R}_2({\cal T}_{99}) = \frac{\sfW_{1234567}+\sft_{34567}}{\sft_{1234}
\sft_{56712}
\sft_{34567}
\sfR_{3456712}
\sfR_{7654321}
\sfR_{2176543}
\sfW_{1234567}}.
\ee
The final answer, ${\cal R}({\cal T}_{99})={\cal R}_1({\cal T}_{99})+{\cal R}_2({\cal T}_{99})$, can be nicely written as the sum over four terms, each with six poles. However, ${\cal R}({\cal T}_{99})$ only has nine distinct poles as expected.

\section{Connection to Cluster Algebras}
\label{sec:clusterconnection}

In 2003 Speyer and Williams introduced the notion of the positive part of tropical Grassmanianns and found deep connections to cluster algebras \cite{SpeyerW}. In a nutshell, cluster algebras consist of a set of variables called cluster variables which form sets called clusters. Each cluster contains the same number of variables and there are exchange rules or mutations that take one cluster to another by mutating a variable \cite{ClusterA,ClusterB}. 

Very recently, Drummond, Foster, G{\"u}rdogan, and Kalousios \cite{Drummond:2019qjk} used the structure of the cluster algebras of $G(3,6)$, $G(3,7)$, and $G(3,8)$ to provide a systematic way of computing biadjoint amplitudes as the volume of the corresponding cluster polytope which is triangulated by the clusters into $2(n-4)-1$ dimensional simplices. 


The underlying object which is triangulated by the cluster algebra is ${\rm Trop}^+G(3,n)$. Given that there is a bijection between metric tree arrangements and facets of ${\rm Trop}\,G(3,n)$ (or more precisely of the Dressian ${\rm Dr}(3,n)$ which contains ${\rm Trop}\,G(3,n)$), it is natural to expect a bijection between planar collections of Feynman diagrams and facets of  ${\rm Trop}^+G(3,n)$ (or more precisely of the positive Dressian ${\rm Dr}^+(3,n)$) \cite{herrmann2009draw}. 


It is know that not all facets of ${\rm Trop}^+G(3,n)$ are simplices. 
For example, as mentioned in section \ref{sec:sevenpexample} in the context of planar collections, for $n=7$ one has $595$ facets with six vertices, $63$ facets with seven vertices, $28$ facets with eight vertices and $7$ facets with $9$ vertices. This gives a total of $833$ simplices, a number which coincides with the number of clusters in the $E_6$ cluster algebra.

From here we conclude that planar collections of trees combine some clusters into single objects. It would be interesting to study ``the algebra acting on planar collections of Feynman diagrams'' which would have a close connection to cluster algebras. 

Let us make some comments on the comparison between the underlying cluster algebra and the one on planar collections: 

\begin{itemize}

\item Each cluster has the same number of cluster variables, i.e. $2(n-4)$, while each planar collection has the same number of trees, i.e. $n$. 

\item Each cluster has as many mutations as the number of variables while planar collections have various numbers of degenerations with $2(n-4)$ as lower bound. From the facet of ${\rm Trop}^+G(3,n)$ (or ${\rm Dr}^+(3,n)$) point of view, this is because facets are not necessarily simplices as mentioned above. 

\item A mutation connects one cluster exactly to one other cluster while each degeneration of a planar collection connects it to exactly one other planar collection.

\end{itemize}

For physical applications, computing the contributions to an amplitude using clusters requires the translation from the algebra to the Grassmannian by producing $2(n-4)$ Pl\"{u}cker vectors. One then performs the inner product of the Pl\"{u}cker vectors with the vector of kinematic invariants \cite{Drummond:2019qjk}. Note that such an inner product coincides with ${\cal F}(\cal T)$ if $2n-9$ degenerations are performed simultaneously while still producing a valid planar collection. Such planar collections correspond to the vertices of the simplices. In our approach, one does not need to determine the Pl\"{u}cker vectors for the vertices. The contribution to an amplitude ${\cal R}({\cal T})$ is computed directly as an integral over the valid metrics of ${\cal T}$. This operation does not require a triagularization but it often requires a separation into regions if an analytic result is desired. A numerical evaluation would not require this last step.

\section{Higher $k$ Objects}
\label{sec:higherk}

We have provided strong evidence for considering planar collections of Feynman diagrams as the natural generalization of standard planar Feynman diagrams. Moreover, computations of (generalized) biadjoint amplitudes for $k=2$ and $k=3$ are now put on equal footing as sums over the corresponding planar objects. Of course, one of the most pressing problems is to find what kind of generalization of quantum field theory leads to planar collections of Feynman diagrams as their perturbative expansion. This is also related to the question of generalizing not only tree but also loop Feynman diagrams. It has been known for a long time that the Feynman tree theorem can relate loop diagrams to trees \cite{Feynman:1963ax, CaronHuot:2010zt}. We leave this fascinating direction for future research and instead we focus here on what objects are needed at tree-level for $k\geq 4$. 

The answer is already suggested by the way we motivated planar collections starting with $n=5$ and using ${\rm Trop}\, G(2,5)\sim {\rm Trop}\, G(3,5)$. When $k=4$, we have to options: start with ${\rm Trop}\, G(2,6)\sim {\rm Trop}\, G(4,6)$ or with ${\rm Trop}\, G(3,7)\sim {\rm Trop}\, G(4,7)$ but both lead to the same answer\footnote{We thank A.Guevara for suggesting the latter as a faster route.}. Let us discuss the former as it also explains the mysterious fact that for every $k=2$ $n$-point Feynman diagram one can construct a planar collection even for $n>5$.

\begin{figure}[t]
    \centering
    \includegraphics[width=\linewidth]{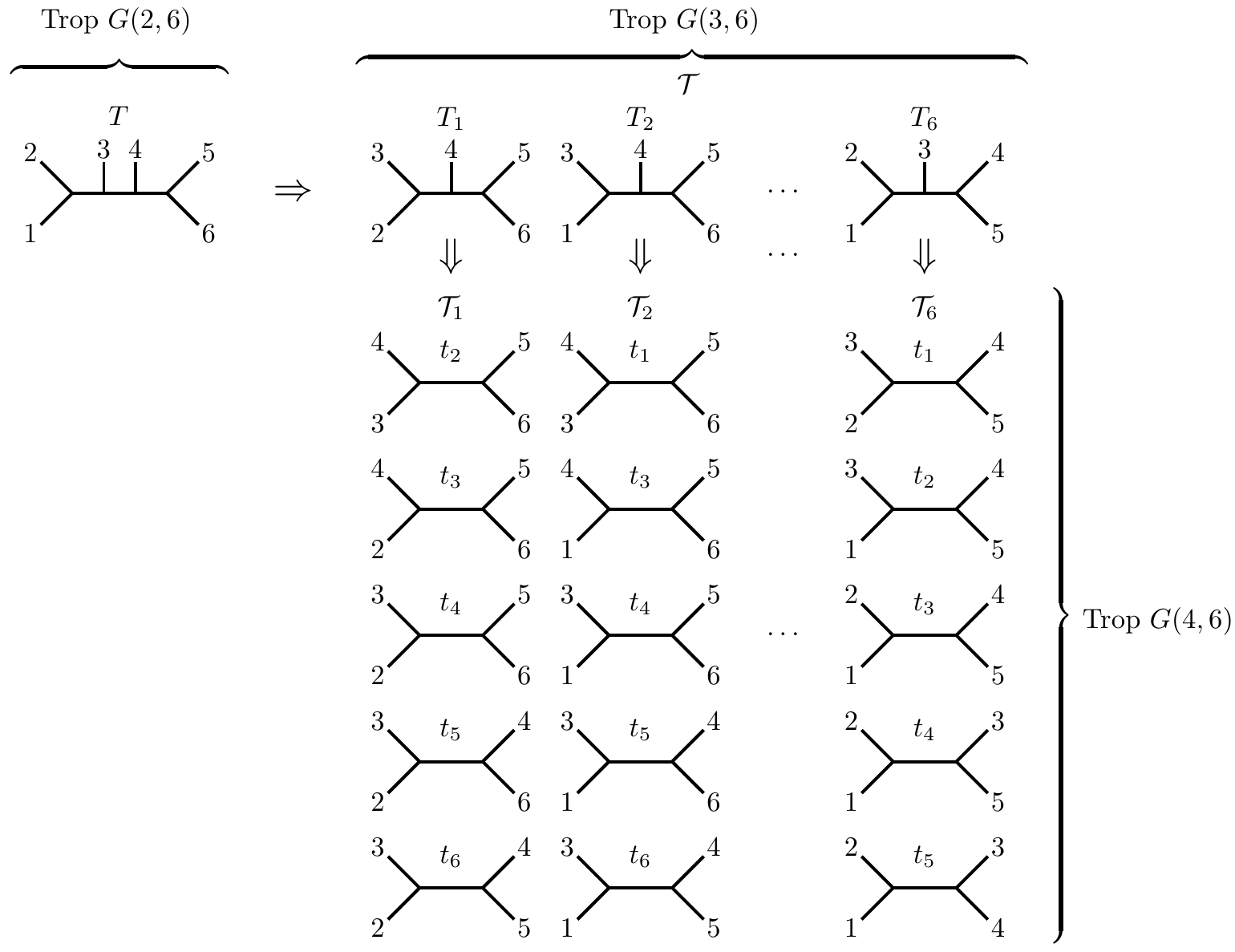}
    \caption{As stated in section \ref{sec:planarfeynman}, one can construct a planar collection $\mathcal{T}$ (related to ${\rm Trop\,}G(3,6)$) from every tree in ${\rm Trop\,}G(2,6)$. Now, from each such planar collection one can generate a planar collection of collections of Feynman diagrams by constructing the dual of each tree in $\mathcal{T}$. This planar collection of collections should contribute to $n=6,k=4$ amplitudes (and be related to ${\rm Trop\,}G(4,6)$). Since ${\rm Trop\,}G(4,6) \sim {\rm Trop\,}G(2,6)$, this ``layered collection'' represents the dual object of the tree $T$.}
    \label{fig:grassmap}
\end{figure}

Consider any $n=6$ planar Feynman diagram $T$ as in figure \ref{fig:grassmap}. As we have shown in section \ref{sec:planarfeynman}, there exists a planar collection associated with each one of them. Each planar collection ${\cal T}$ has six trees with five leaves. Now, each five-point planar tree is strictly dual to a planar collection of five trees with four leaves. We propose that the correct object for $k=4$ is nothing but the corresponding planar collection of collections!

It must be that the $14$ planar collections of collection generated in this way satisfy the natural generalization of the compatibility conditions for metrics $d^{ij}_{ab}$, i.e. that $\pi_{abef}:=d^{eg}_{ab}$ is a completely symmetric rank-four tensor. Moreover, such $14$ objects should always have only three degenerations as the dual $n=6$ Feynman diagram have exactly three degenerations. 

Note that, still in the context of $n=6$ and $k=4$, if the construction is correct, one should find that any of the $48$ planar collections for $n=6$ and $k=3$ which is not in the set of $14$ special ones, leads to an object that does not admit a metric satisfying the compatibility condition.  

We expect that if $n\geq 7$ then every single collection for $k=3$ would lead to a valid collection of collections in $k=4$ via the construction defined above. 

Clearly, the procedure can be continued to higher values of $k$ but this is out of the scope of this paper and we leave a detailed study of $k>3$ to future work.

\section*{Acknowledgements}

We would like to thank A. Guevara for very useful discussion and N. Early for many discussions on matroid subdivisions of the hypersimplex and relations to tree arrangements. We also would like to thank both of them for comments on the manuscript. Research at Perimeter Institute is supported in part by the Government of Canada through the Department of Innovation, Science and Economic Development Canada and by the Province of Ontario through the Ministry of Economic Development, Job Creation and Trade.

\appendix

\section{Mathematical Background on Metric Tree Arrangements}
\label{ap:treearrs}

In this appendix we review the definition of abstract tree arrangements, metric tree arrangements, and the relation to ${\rm Trop}\, G(3,n)$. In this review we follow very closely the discussion by Herrmann, Jensen, Joswig, and Sturmfels \cite{herrmann2009draw} and refer the reader to their paper for a much more in depth discussion.

In \cite{herrmann2009draw} planarity is not used as a constraint so in this appendix we also drop it and discuss completely general tree arrangements. 

As we have seen in the main text, a metric tree arrangement is a very constrained object due to the compatibility on the metrics. A purely combinatorical test to decide whether a set of trees admits a metric is not known. However, a necessary condition, which is also sufficient for $n<9$, is the condition to be an abstract tree arrangement.

\begin{definition}
\cite{herrmann2009draw} An $n$-tuple of trees ${\cal T}=(T_1,T_2,\ldots ,T_n)$ with $T_i$ a tree on $[n]\setminus i$ is an {\it abstract tree arrangement} if
\begin{itemize}
    \item either $n=4$;
    \item or $n=5$ and ${\cal T}$ is obtained from a single tree with $5$ leaves by pruning on leaf at a time;
    \item or $n>5$ and $(T_1\setminus i,T_2\setminus i,\ldots ,T_{i-1}\setminus i,T_{i+1}\setminus i,\ldots ,T_n\setminus i)$ is an abstract tree arrangement for all $i\in [n]$. Here $T_m\setminus i$ means the tree obtained by removing leaf $i$ from $T_m$.
\end{itemize}

\end{definition}

Note that this definition is recursive and has a beautiful interpretation as a soft-limit check. Removing a given leaf from all trees that contain it is the physical operation of taking a particle soft. 

Unfortunately, this test is not sufficient for $n\geq 9$ as the following example, also given in \cite{herrmann2009draw} demonstrates:
\be\label{nonP}
\begin{gathered}
\{ C(24,6598,37)\qc C(14,5768,39) \qc C(17,5846,29)\qc C(12,6579,38)\qc C(26,4198,37),\\ C(14,5729,38)\qc C(13,5894,26)\qc C(15,7346,29)\qc C(15,7468,23)\}.
\end{gathered}
\ee
These tree are all caterpillars and e.g. $C(24,6598,37)$ has leaves $2$ and $4$ on the left cherries, $3$ and $7$ on the right cherries while $6,5,9,8$ are the legs in that order from left to right. This is an abstract tree arrangement but it does not admit a set of compatible metrics and therefore it is not a metric tree arrangement.

We have checked that this abstract tree arrangement is not planar. In other words, there is no permutation of $\{1,2,\ldots ,9\}$ such that each one of the nine trees in \eqref{nonP} is planar with respect to the order induced by the permutation after removing the $i^{\rm th}$ label.

This leads us to ask the following:

\noindent {\bf Question A.2}. 
Do all planar abstract arrangements of trees admit a set of compatible metrics? Or in other words, is the set of all planar abstract tree arrangements the same as the set of all planar metric tree arrangements?

Finally, we comment on the connection between metric tree arrangements and tropical geometry. 

Given any metric tree arrangement, ${\cal T}$, the compatibility conditions on the metrics implies that $d^{(a)}_{bc}$ can be turned into a completely symmetric tensor, i.e., $\pi_{abc}:=d^{(a)}_{bc}$. The claim is that $\pi_{abc}$ is a Pl\"{u}cker vector of the Dressian ${\rm Dr}(3,n)$. The Dressian is a tropical space closely related to the tropical Grassmannian. In fact, ${\rm Tr}\,G(3,n)$ sits inside ${\rm Dr}(3,n)$. The reason is that ${\rm Dr}(3,n)$ is defined as the space of vectors that satisfy the three-term tropical Pl\"{u}cker relations while vectors in ${\rm Tr}\,G(3,n)$ are required to also satisfy all tropical Pl\"{u}cker relations with four of more terms.

In order to see that $\pi_{abc}$ defines a ray of the Dressian one has to slighly modify the definition when any two indices coincide, i.e. $\pi_{aab}:=\infty$. Once this is done it is not difficult to show that the three-term tropical Pl\"{u}cker relations are satisfied.

Let us recall that in tropical geometry, a tropical variety is not defined by the zeros of a set of polynomials as in standard algebraic geometry. Instead, after tropicalizing the set of polynomial, the variety is defined by points where at least two of the linear functions in the tropical polynomial achieve the minimum value in the set.

The standard three-term tropical Pl\"{u}cker relations read
\be
{\cal V}_{ij,kl}^h:=\pi_{hij}\pi_{hkl}-\pi_{hik}\pi_{hjl}+\pi_{hil}+\pi_{hjk} = 0.
\ee
while the tropical version of ${\cal V}_{ij,kl}^h$ becomes
\be\label{plus}
{\rm min}\,\{ \pi_{hij}+\pi_{hkl}, \pi_{hik}+\pi_{hjl}, \pi_{hil}+\pi_{hjk}\}.
\ee
As mentioned above, the tropical version of requiring ${\cal V}_{ij,kl}^h=0$ is asking for points where \eqref{plus} achieves a minimum at least twice. More explicitly, one has to require that either
\begin{equation}
\begin{aligned}
\pi_{hij}+\pi_{hkl} & =  \pi_{hik}+\pi_{hjl}<\pi_{hil}+\pi_{hjk}\quad {\rm or}\\
\pi_{hij}+\pi_{hkl} & =  \pi_{hil}+\pi_{hjk}<\pi_{hik}+\pi_{hjl}\quad {\rm or}\\
\pi_{hik}+\pi_{hjl} & =  \pi_{hil}+\pi_{hjk}<\pi_{hij}+\pi_{hkl}.
\end{aligned}
\end{equation}

\bibliographystyle{JHEP}
\bibliography{references}

\providecommand{\href}[2]{#2}\begingroup\raggedright\begin{thebibliography}{10}

\bibitem{Cachazo:2013iea}
F.~Cachazo, S.~He, and E.~Y. Yuan, {\it {Scattering of Massless Particles:
  Scalars, Gluons and Gravitons}},  {\em JHEP} {\bf 07} (2014) 033,
  [\href{http://arxiv.org/abs/1309.0885}{{\tt arXiv:1309.0885}}].

\bibitem{Strassler:1992zr}
M.~J. Strassler, {\it {Field theory without Feynman diagrams: One loop
  effective actions}},  {\em Nucl. Phys.} {\bf B385} (1992) 145--184,
  [\href{http://arxiv.org/abs/hep-ph/9205205}{{\tt hep-ph/9205205}}].

\bibitem{BilleraL}
L.~J. Billera, S.~P. Holmes, and K.~Vogtmann, {\it Geometry of the space of
  phylogenetic trees},  {\em Adv. Appl. Math.} {\bf 27} (Nov., 2001) 733--767.

\bibitem{herrmann2009draw}
S.~Herrmann, A.~Jensen, M.~Joswig, and B.~Sturmfels, {\it How to draw tropical
  planes},  {\em the electronic journal of combinatorics} {\bf 16} (2009),
  no.~2 6.

\bibitem{Cachazo:2019ngv}
F.~Cachazo, N.~Early, A.~Guevara, and S.~Mizera, {\it {Scattering Equations:
  From Projective Spaces to Tropical Grassmannians}},
  \href{http://arxiv.org/abs/1903.08904}{{\tt arXiv:1903.08904}}.

\bibitem{Cachazo:2019apa}
F.~Cachazo and J.~M. Rojas, {\it {Notes on Biadjoint Amplitudes, ${\rm
  Trop}\,G(3,7)$ and $X(3,7)$ Scattering Equations}},
  \href{http://arxiv.org/abs/1906.05979}{{\tt arXiv:1906.05979}}.

\bibitem{Drummond:2019qjk}
J.~Drummond, J.~Foster, {\"O}.~G{\"u}rdogan, and C.~Kalousios, {\it {Tropical
  Grassmannians, cluster algebras and scattering amplitudes}},
  \href{http://arxiv.org/abs/1907.01053}{{\tt arXiv:1907.01053}}.

\bibitem{Sepulveda:2019vrz}
D.~García~Sepúlveda and A.~Guevara, {\it {A Soft Theorem for the Tropical
  Grassmannian}},  \href{http://arxiv.org/abs/1909.05291}{{\tt
  arXiv:1909.05291}}.

\bibitem{speyer2004tropical}
D.~Speyer and B.~Sturmfels, {\it The tropical grassmannian},  {\em Advances in
  Geometry} {\bf 4} (2004), no.~3 389--411.

\bibitem{speyer2004tropicalM}
D.~Speyer and B.~Sturmfels, {\it Tropical mathematics},  {\em arXiv preprint
  math/0408099} (2004).

\bibitem{Cachazo:2013hca}
F.~Cachazo, S.~He, and E.~Y. Yuan, {\it {Scattering of Massless Particles in
  Arbitrary Dimensions}},  {\em Phys. Rev. Lett.} {\bf 113} (2014), no.~17
  171601, [\href{http://arxiv.org/abs/1307.2199}{{\tt arXiv:1307.2199}}].

\bibitem{Cachazo:2014xea}
F.~Cachazo, S.~He, and E.~Y. Yuan, {\it {Scattering Equations and Matrices:
  From Einstein To Yang-Mills, DBI and NLSM}},  {\em JHEP} {\bf 07} (2015) 149,
  [\href{http://arxiv.org/abs/1412.3479}{{\tt arXiv:1412.3479}}].

\bibitem{SpeyerW}
D.~{Speyer} and L.~K. {Williams}, {\it {The tropical totally positive
  Grassmannian}},  {\em arXiv Mathematics e-prints} (Dec, 2003) math/0312297,
  [\href{http://arxiv.org/abs/math/0312297}{{\tt math/0312297}}].

\bibitem{ClusterA}
S.~{Fomin} and A.~{Zelevinsky}, {\it {Cluster algebras I: Foundations}},  {\em
  arXiv Mathematics e-prints} (Apr, 2001) math/0104151,
  [\href{http://arxiv.org/abs/math/0104151}{{\tt math/0104151}}].

\bibitem{ClusterB}
S.~{Fomin} and A.~{Zelevinsky}, {\it {Cluster algebras II: Finite type
  classification}},  {\em Inventiones Mathematicae} {\bf 154} (Oct, 2003)
  63--121, [\href{http://arxiv.org/abs/math/0208229}{{\tt math/0208229}}].

\bibitem{Feynman:1963ax}
R.~P. Feynman, {\it {Quantum theory of gravitation}},  {\em Acta Phys. Polon.}
  {\bf 24} (1963) 697--722. [272(1963)].

\bibitem{CaronHuot:2010zt}
S.~Caron-Huot, {\it {Loops and trees}},  {\em JHEP} {\bf 05} (2011) 080,
  [\href{http://arxiv.org/abs/1007.3224}{{\tt arXiv:1007.3224}}].

\end{thebibliography}\endgroup

\end{document}